\documentclass[]{jfm}
\usepackage{graphicx}
\usepackage{newtxtext}
\usepackage{newtxmath}
\usepackage{natbib}
\usepackage{hyperref}
\usepackage{float}
\usepackage{subfig}
\usepackage{dcolumn}
\usepackage{bm}
\usepackage{gensymb}
\usepackage{tabularx}
\usepackage{color}
\usepackage{braket}
\usepackage{comment}
\newcommand{\RomanNumeralCaps}[1]
    {\MakeUppercase{\romannumeral #1}}
\hypersetup{colorlinks = true,
	    urlcolor   = blue,
	    citecolor  = blue,}

\title{Natural convection in a vertical channel. Part 1. Wavenumber interaction and Eckhaus instability in a narrow domain}
\shorttitle{Natural convection in a vertical channel. Part 1}
\shortauthor{Z. Zheng, L.S. Tuckerman and T.M. Schneider}

\author{Zheng Zheng\aff{1},
	Laurette S. Tuckerman\aff{2} 
        \corresp{\email{laurette.tuckerman@espci.fr}}
        \and Tobias M. Schneider\aff{1}}

\affiliation{
    \aff{1} Emergent Complexity in Physical Systems Laboratory (ECPS), \'Ecole Polytechnique F\'ed\'erale de Lausanne, CH 1015 Lausanne, Switzerland
    \aff{2} Physique et M\'ecanique des Milieux H\'et\'erog\`enes (PMMH), CNRS, ESPCI Paris, PSL University, Sorbonne Universit\'e, Universit\'e de Paris, 75005 Paris, France}
 
\begin{document}
\maketitle

\begin{abstract}
In a vertical channel driven by an imposed horizontal temperature gradient, numerical simulations \citep{Gao2013, Gao2015, Gao2018} have previously shown steady, time-periodic and chaotic dynamics. We explore the observed dynamics by constructing invariant solutions of the three-dimensional Oberbeck–Boussinesq equations, characterizing the stability of these equilibria and periodic orbits, and following the bifurcation structure of the solution branches under parametric continuation in Rayleigh number. We find that in a narrow vertically-periodic domain of aspect ratio ten, the flow is dominated by the competition between three and four co-rotating rolls. We demonstrate that branches of three and four-roll equilibria are connected and can be understood in terms of their discrete symmetries. Specifically, the $D_4$ symmetry of the four-roll branch dictates the existence of qualitatively different intermediate branches that themselves connect to the three-roll branch in a transcritical bifurcation due to $D_3$ symmetry. The physical appearance, disappearance, merging and splitting of rolls along the connecting branch provide a physical and phenomenological illustration of the equivariant theory of $D_3$-$D_4$ mode interaction. We observe other manifestations of the competition between three and four rolls, in which the symmetry in time or in the transverse direction is broken, leading to limit cycles or wavy rolls, respectively. Our work highlights the interest of combining numerical simulations, bifurcation theory, and group theory, in order to understand the transitions between and origin of flow patterns.
\end{abstract}


\section{Introduction}
\par A fluid subjected to a horizontal temperature gradient, often called natural or vertical convection, is encountered in a wide range of geophysical \citep{Hart1971a}, meteorological and engineering \citep{Kaushika2003, Arici2015} applications. Scientific research on natural convection with its many variants has a long history. Motivated by the crucial application of the thermal insulation, \citet{batchelor1954heat} sought to determine the width which maximized the insulating properties of an air-filled cavity within a wall or window, i.e.\ the double-glazing problem; this solution was later amended by \cite{poots1958heat} and \cite{gershuni1966numerical}. \cite{elder1965laminar} observed, both experimentally and theoretically, the oblique convection rolls that form in a tall enclosure that we will study here. These rolls, in particular their onset, were further investigated by, e.g.\, \cite{eckert1961natural, vest1969stability, gershunibook, korpela1973stability, mizushima1976stability}. After nonlinear numerical simulations became feasible, a number of researchers studied the evolution of the number of rolls in an air-filled cavity with height-to-width aspect ratio eight to twenty by means of time integration \citep{roux1980reverse, le1990note, wakitani1998flow} or by Newton-Krylov methods \citep{salinger2002computational, mizushima2002transitions, gelfgat2004stability, xin2006natural}. Among the phenomena they observed were hysteresis, multiplicity of solutions, and a non-monotonic evolution in the number of rolls with Rayleigh number. 

\par Attesting to its importance, natural convection has been chosen as a computational benchmark for evaluating the accuracy and efficiency of fluid dynamics codes. The benchmark competition on an air-filled square cavity \citep{de1983natural} attracted thirty-seven contributions, comparing results from codes described in, for example, \citet{gil1981influence, winters1982predictions, le1985computation, winters1987hopf, le1991accurate}. An entire conference and journal volume were devoted to the benchmark problem of an air-filled height-to-width-ratio of eight \citep{christon2002computational}.

\par Continuing our survey of archetypal configurations in vertical convection, low-Prandtl-number liquid metals are used in the process of growing semiconductor crystals; the goal is to avoid transition to oscillatory flow that engenders imperfections. A shallow cavity of height-to-width-ratio of 1/4 filled with a low-Prandtl-number liquid metal was the topic of yet another benchmark \citep{roux1990numerical}. Bifurcation analyses of this configuration have been carried out by, e.g., \cite{winters1988oscillatory, pulicani1990spectral, henry1998two, gelfgat1999stability}. We will continue our survey of the literature in \cite{Zheng2023part2}, where we will focus on three-dimensional patterns. 

\par Vertical convection is a special case of inclined layer convection, in which the container is tilted against gravity so that both buoyancy and shear forces drive the flow \citep{poots1958heat, fujimura1993mixed, Daniels2000, Subramanian2016}. Extrapolating in inclination angle from the well-understood buoyancy-driven Rayleigh-B\'enard case to shear-dominated vertical convection may give insights into transition in pure shear flows such as plane Couette flow \citep{nagata1983three}. This was one of the motivations for the extensive study of inclined layer convection by \citet{Reetz2020a, Reetz2020b}, whose spirit and methods are carried over to the present study.

\par Rayleigh-B\'enard convection, in which the layer is horizontal, is probably the most studied case of inclined layer convection, but it is exceptional in several important respects: the Prandtl number $Pr$ (ratio of kinematic viscosity to thermal diffusivity) plays no role at threshold, and the primary instability is always steady. In contrast, \cite{korpela1973stability} showed that the primary instability in vertical convection is steady for $Pr<12.7$ and oscillatory for $Pr>12.7$, a value that was refined to $Pr=12.45$ by \cite{fujimura1993mixed}.

\par Rayleigh-B\'enard convection is also exceptional in that its basic state is motionless, so that lateral boundaries can be assumed to affect only the regions immediately adjacent to them. The interior of a finite domain can therefore be approximated as homogeneous in the horizontal directions parallel to the rigid boundaries and so periodic boundary conditions can be used in these directions. In contrast, in vertical convection, the basic state includes a velocity which is vertical and hence normal to boundaries situated at the top and bottom. Such boundaries can have a substantial influence on the basic solution in the bulk if the aspect ratio is low or the Rayleigh number is high. This undermines the approximation of vertical homogeneity, without which theoretical or numerical treatment becomes much more difficult. \cite{batchelor1954heat, eckert1961natural, gill1966boundary,vest1969stability, mizushima1976stability, bergholz1978instability} distinguished two regimes for the basic solution in the bulk of a finite cavity: the conductive regime, in which the temperature depends only on the distance from the walls, and the double boundary layer regime, in which the temperature also has a vertical gradient resulting from the flow meeting the upper and lower boundaries. The researchers cited above showed that even in the boundary layer regime, a cavity of finite height can be approximated by a vertically homogeneous problem if modified boundary conditions are imposed on the temperature at the two vertical bounding plates, either a finite vertical gradient or else horizontal isoflux conditions \citep{kimura1984boundary, le2022natural}. The configuration that we study here, with an aspect ratio of ten and Rayleigh numbers lower than $14\ 000$, falls safely into the conductive regime. This means that our simulations using periodic vertical boundary conditions and bounding plates each of constant temperature resemble the flow in the interior regions of cavities of finite height. For a full treatment of the differences between periodic domains and those with boundaries (free-slip or rigid), see \cite{hirschberg1997mode}. 

\par Our investigation is based on a series of studies by \cite{Gao2013, Gao2015, Gao2018}. These authors used direct numerical simulations (DNS) combined with linear and weakly non-linear approaches to study vertical convection in air ($Pr=0.71$). By systematically increasing the Rayleigh number, \citet{Gao2013} surveyed the regimes in a three-dimensional domain whose periodic vertical height was ten times that of the other two. They observed that the flow transitioned from the conductive state to steady rolls, then to oscillatory flow, and finally to a chaotic state. After acquiring four identical stacked co-rotating rolls, the flow continued to have a vertical periodicity of a quarter of the domain length over a fairly large Rayleigh-number range. By subsequently confining the domain to this height to suppress large-scale instabilities, \citet{Gao2015} observed a period-doubling cascade leading to chaotic dynamics as the Rayleigh number was increased. However, a quantitative numerical bifurcation analysis corresponding to these studies has not been performed and thus the bifurcation-theoretic origins of the observed complex flow patterns remain to be fully explored. This motivates the present study. 

\par
\begin{figure}
    \centering
    \includegraphics[width=\columnwidth]{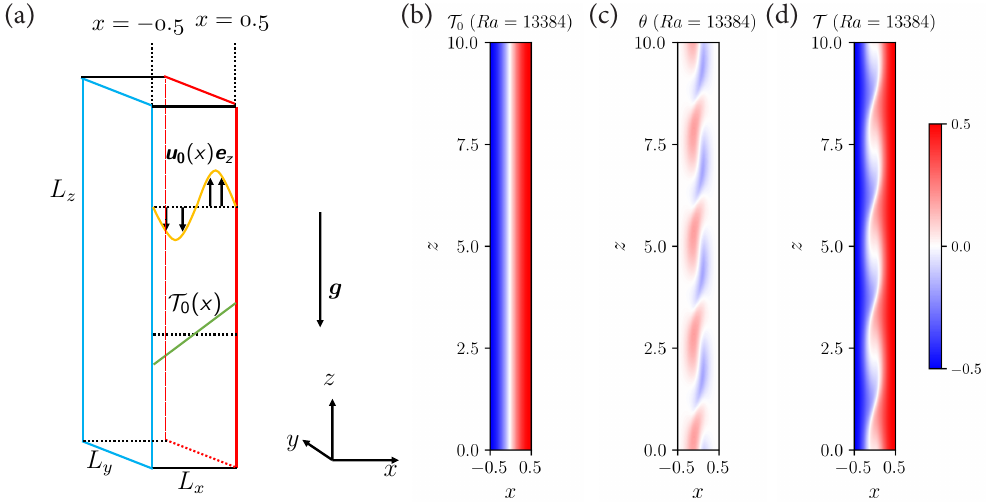}
    \captionsetup{font={footnotesize}}
    \captionsetup{width=13.5cm}
    \captionsetup{format=plain, justification=justified}
    \caption{\label{VC_figure} (a) Schematic of the computational domain. A fluid layer is bounded between two walls located at $x=\pm0.5$. The temperature at wall $x=0.5$ is fixed to a higher value than that at wall $x=-0.5$. The long $z$ direction is aligned with gravity; both the $y$ and $z$ directions are taken to be periodic with spatial periods $L_y=1$ and $L_z=10$. The orange curve and green line show the cubic velocity \eqref{conductivea} and linear temperature \eqref{conductiveb} profiles of the conductive base solution. (b)-(d) Temperature $\mathcal{T}_0$ of the basic state, temperature deviation $\theta \equiv \mathcal{T} - \mathcal{T}_0$ and total temperature field $\mathcal{T}$ of the convection roll structure (FP1 in figure \ref{small_FP}) visualized in the $x$-$z$ plane on the arbitrary plane $y=0.5$ at $Ra=13384$.}
\end{figure}

 We consider the domain $[L_x,L_y,L_z]=[1,1,10]$, where $x$, $y$ and $z$ represent the direction between the two bounding plates, the transverse direction, and the direction of gravity, respectively, shown in figure \ref{VC_figure}. Here, only one of the three spatial directions is extended and thus the resulting flow structures, while fascinating and surprisingly complex, predominantly vary only in the vertical direction. Although the domain has only one spatially extended direction, weakly two-dimensional patterns have also been observed. Note, however, that all computations of \cite{Gao2013, Gao2015, Gao2018} as well as those presented here are fully three-dimensional. The domain $[L_x,L_y,L_z]=[1,8,9]$ studied by \cite{Gao2018} has two extended directions and correspondingly fully two-dimensional patterns and behavior. A bifurcation-theoretic analysis of these will be the subject of our companion paper \citet{Zheng2023part2}.

\par Above onset, as the Rayleigh number is increased, a sequence of convective patterns emerges from the conductive state. At each bifurcation point, symmetries of the previous state are in general sequentially broken, leading to patterns of increasing complexity. Those sequentially broken symmetries include continuous or $n$-fold translation symmetry, reflection symmetry, centro-symmetry and so on. The transition to $n$-fold translation symmetry in an effectively one-dimensional and reflection-symmetric domain generically leads to $D_n$ symmetry. The phenomenon of competition between steady branches with different wavenumbers is the essence of the Eckhaus instability \citep{eckhaus1965studies, tuckerman1990bifurcation}, especially in the idealized case of long domains. For the particular finite vertical aspect ratio of ten in our convection problem, four co-rotating rolls are favored, competing with three rolls at increasing $Ra$. As it happens, symmetry groups $D_4$ and $D_3$ have very particular properties; it is this combination of group theory, topology, and fluid mechanics which shape the resulting bifurcation diagram. The competition between three and four rolls is also manifested by branches of time-dependent states in which the number of rolls alternates periodically or chaotically. 

\par More generally, \cite{dangelmayr1986steady} carried out a comprehensive investigation using weakly nonlinear model equations of the scenarios resulting from the competition between periodic patterns with different wavenumbers. \cite{crawford1990period} applied similar equations to a Faraday wave experiment. Among the features of these scenarios are steady pure-mode (single wavenumber) and mixed-mode (multiple wavenumber) branches, as well as traveling and standing waves. One of the main topics of our investigation is the numerical simulation and qualitative interpretation of the mixed-mode branches in our hydrodynamic system.

\par We begin by reproducing the equilibria and periodic orbits computed by \citet{Gao2013}. By following the branches to which these solutions belong, we discover new solution branches and identify the bifurcations giving rise to all of them, from onset to the chaotic regime. The remainder of the manuscript is structured as follows: in \S 2 we discuss the governing equations, numerical aspects, symmetries, and the measurements and visualizations to be presented. \S 3 will present the steady solutions or equilibria, together with a detailed interpretation of the observed bifurcation scenarios using $D_4$ and $D_3$ symmetry \citep{swift1985bifurcation, gambaudo1985perturbation, Knobloch1986, golubitsky1988singularities, Dawes2005}. Time-periodic solutions will be presented in \S 4. Finally, we conclude by a summary of key results and a discussion in \S 5.

\section{Vertical convection system and numerical aspects}

\subsection{Governing equations}
\label{governing-equations}
\par We used the ILC extension module of the MPI-parallel pseudo-spectral code \textit{Channelflow 2.0} \citep{ReetzPhD, Gibson2019} to carry out direct numerical simulations (DNS) of the non-dimensionalized Oberbeck–Boussinesq equations 
\begin{subequations}
\label{appabc}
\begin{align}
    \dfrac{\partial \boldsymbol u}{\partial t} + (\boldsymbol u \cdot \nabla)\boldsymbol u &= -\nabla p + \sqrt{\frac{Pr}{Ra}} \nabla^2 \boldsymbol u + \mathcal{T}\boldsymbol e_z, \label{appa} \\
    \dfrac{\partial \mathcal{T}}{\partial t} + (\boldsymbol u \cdot \nabla)\mathcal{T} &= \frac{1}{\sqrt{Pr\: Ra}}\nabla^2 \mathcal{T}, \label{appb} \\
    \nabla \cdot \boldsymbol u&= 0,	\label{appc} 
\end{align}
\end{subequations}
in a vertical channel depicted in figure \ref{VC_figure}. In equations \eqref{appabc}, $\boldsymbol u = [u, v, w](x,y,z,t)$ and $\mathcal{T}=\mathcal{T}(x,y,z,t)$ stand for total velocity and temperature respectively. The constant buoyancy term has been omitted from \eqref{appa}; correspondingly, the pressure $p=p(x,y,z,t)$ is relative to the hydrostatic pressure. Bold symbols denote vector quantities and $\boldsymbol e_z$ is the vertical unit vector. The equations have been non-dimensionalized with respect to the temperature difference $\Delta \vartheta$ and the distance $W$ between the walls, and the free-fall time unit $(W/g\alpha \Delta \vartheta)^{1/2}$ where $\alpha$ is the thermal expansion coefficient and $g$ is the gravitational acceleration. Two independent dimensionless parameters appear: Rayleigh number $Ra = g \alpha \Delta \vartheta W^3/(\nu \kappa)$ and Prandtl number $Pr = \nu/\kappa$, where $\nu$ is the kinematic viscosity and $\kappa$ the thermal diffusivity. 

\par Periodic boundary conditions are imposed in the $y$ and $z$ directions with spatial periods of $L_y$ and $L_z$, respectively. The walls are no-slip and have prescribed temperatures:  
\begin{align}
    \boldsymbol u(x=\pm 0.5) = 0,
    \qquad 
    \mathcal{T}(x=\pm0.5)=\pm0.5,
\label{eq:bcs}
\end{align}
A supplementary integral constraint on either the pressure gradient or mean flux must be set in the periodic directions. In order to match the simulations of \citet{Gao2013, Gao2015, Gao2018}, we impose a mean pressure gradient of zero in $y$ and in $z$. Equations \eqref{appabc} together with the boundary conditions admit the conductive solution sketched in figure \ref{VC_figure}(a):
\begin{subequations}
\label{conductiveabc}
\begin{align}
    &\boldsymbol u_0(x) = \frac{1}{6}\sqrt{\frac{Ra}{Pr}} \left(\frac{1}{4}x - x^3 \right) \boldsymbol e_z,	\label{conductivea} \\
    &\mathcal{T}_0(x) = x, \label{conductiveb} \\
   &p_0(x) = \Pi,	\label{conductivec} 
\end{align}
\end{subequations}
with arbitrary pressure constant $\Pi$.

\subsection{Numerical methods}
\par Channelflow-ILC adopts Chebychev-Fourier–Fourier (in $x$, $y$ and $z$) expansions for representing flow fields in space and finite differencing method for time integration (see detailed description in \emph{Appendix A} of \citet{Reetz2020a}). We have simulated the three-dimensional computational domain studied in \cite{Gao2013}. This narrow domain $[L_x, L_y, L_z] = [1, 1, 10]$ is discretized by $[N_x, N_y, N_z] = [31, 32, 96]$ collocation points, resulting in a state space dimension $N=4 \times N_x \times N_y \times N_z \times (\frac{2}{3})^2$ on the order of $2\times10^5$. The factor four stems from three components of velocity field and one in temperature field, and $(2/3)^2$ is due to dealiasing in two Fourier directions \citep{Canuto2006}. Although our resolution is slightly less than that reported in \citet{Gao2013}, we find it to be sufficient, since the ratio of the $L_2$-norm of the last resolved mode to the first mode of the velocity and temperature fields is less than $10^{-6}$ in the $y$- and $z$-directions and less than $10^{-9}$ in the $x$-direction, a criterion also employed by \citet{Gibson2016homoclinic}.

\par As an extension to the studies based on DNS observations \citep{Gao2013, Gao2015, Gao2018}, our objective is to construct the invariant solutions such as equilibria and periodic orbits underlying the complex spatio-temporal flow dynamics. For identifying linearly stable states, time-marching (DNS) appropriate initial conditions gives access to these solutions, which is how \citet{Gao2013, Gao2015, Gao2018} proceeded. However, the root-finding technique is required for constructing unstable states. Invariant solutions are state vectors $\boldsymbol{x}^{*}(t)$ satisfying 
\begin{equation}
    \mathcal{G}(\boldsymbol{x}^{*})=\sigma \mathcal{F}^T(\boldsymbol{x}^{*}) - \boldsymbol{x}^{*} = 0,
    \label{invariant_equation}
\end{equation}
where $\sigma$ is a symmetry operator and $\mathcal{F}^T$ is the time-evolution operator integrating \eqref{appa} – \eqref{appc} from an initial state $\boldsymbol{x^*}$ over a finite time period $T$ (where $T$ is the period of a periodic solution, and arbitrary for a steady solution). The shooting-based Newton–Raphson method in Channelflow-ILC uses a matrix-free Krylov method in which successive Krylov vectors are generated by time-marching initial conditions \citep{Kelley2003, Sanchez2004}. It is usually combined with a hookstep trust-region optimization based on the Krylov vectors, leading to a greatly increased radius of convergence \citep{Viswanath2007, Viswanath2009}. Convergence is considered to be reached once the norm of the residual of \eqref{invariant_equation} is sufficiently close to machine precision (of the order of $10^{-12}$). The converged solutions are subsequently continued parametrically along a range of Rayleigh numbers to form bifurcation diagrams \citep{Sanchez2004, Dijkstra2014} so as to understand their bifurcation structure. 

\par The stability of each converged state is evaluated by using the Arnoldi algorithm \citep{Arnoldi1951, Antoulas2005} to determine its leading eigenvalues and eigenvectors for fixed points, or Floquet exponents and Floquet modes for periodic orbits. In a highly symmetric problem like this one, most eigenvalues are multiple, since symmetry operations applied to non-symmetric eigenvectors can yield other eigenvectors. For multiple eigenvalues, the Arnoldi algorithm returns an arbitrary set of linearly independent eigenvectors. We take linear combinations of these to construct those eigenvectors within the respective eigenspaces that are appropriate for our purposes. 

\subsection{Symmetries of the system}
\par The vertical convection system is equivariant under $y$-reflection \eqref{sym_a}, combined $x$ and $z$ reflection \eqref{sym_b}, and translation in $y$ and $z$ \eqref{sym_c}:
\begin{subequations}
    \label{sym_all}
    \begin{eqnarray}
        &\pi_y[u,v,w,\mathcal{T}](x,y,z) \equiv [u,-v,w,\mathcal{T}](x, -y,z) \label{sym_a},\\
        &\pi_{xz}[u,v,w,\mathcal{T}](x,y,z) \equiv [-u,v,-w,-\mathcal{T}](-x, y,-z) \label{sym_b}, \\
        &\tau(\Delta y, \Delta z)[u,v,w,\mathcal{T}](x,y,z) \equiv [u,v,w,\mathcal{T}](x, y + \Delta y,z + \Delta z). \label{sym_c}
    \end{eqnarray}
\end{subequations}
Since $y$ and $z$ are periodic directions, the center of reflection $(y_0,z_0)$ is arbitrary, and so reflections $y\rightarrow -y$ and $z\rightarrow -z$ in \eqref{sym_a} and \eqref{sym_b} should be more generally written as $y_0+y\rightarrow y_0-y$ and $z_0+z\rightarrow z_0-z$ for some $y_0$ and $z_0$. We will write these merely as $\pi_y$ and $\pi_{xz}$, while for visualizations we will choose whatever axis of reflection seems most appropriate for $y_0$ and $z_0$, usually $L_y/2$ and $L_z/2$. 

\par The symmetry transformations \eqref{sym_all} form the equivariance group of the system, which consists of all products of the generators $S_{VC} \equiv \braket{\pi_y, \pi_{xz}, \tau(\Delta y, \Delta z)}\simeq [O(2)]_y \times [O(2)]_{xz}$. (Although symmetry groups cannot always be associated only with $y$ or with $(x,z)$, we will do so occasionally when this is convenient and possible.) The groups that arise in this study are $Z_n$, the cyclic group of $n$ elements, $D_n$, the cyclic group of $n$ elements together with a non-commuting reflection, and $O(2)$, the group of all rotations (or equivalently translations in our periodic domain) together with a non-commuting reflection. We note that $D_1 =Z_2$ and $D_2=Z_2\times Z_2$. Aside from the conductive solution, which is invariant under the full group $S_{VC}$, other solutions may be invariant only under proper (smaller) subgroups of $S_{VC}$. Trajectories that begin in an invariant subspace remain so under exact arithmetic, but may depart due to instability. At times in this study, we have imposed reflection symmetries or periodicity over an interval shorter than $L_y$ or $L_z$ in order to restrict the dynamics to the desired invariant subspace or to expedite numerical continuation.

\subsection{Numerical measurements and visualisations}
\par We define the deviation from the conductive solution $\theta \equiv \mathcal{T} - \mathcal{T}_0$, which we shall usually refer to merely as the temperature, and employ its $L_2$-norm
\begin{equation}
    \lvert\lvert \theta \lvert\lvert_2 = \left(\frac{1}{L_y}\frac{1}{L_z}\int_{-0.5}^{0.5}\int_{0}^{L_y} \int_{0}^{L_z} \theta^2(x, y, z)\, dxdydz \right)^{\frac{1}{2}},
    \label{energy}
\end{equation}
as an observable for plotting the bifurcation diagrams. For fixed points (FPs), a single curve representing $\lvert\lvert \theta \lvert\lvert_2$ as a function of the Rayleigh number is plotted. For periodic orbits (POs), the maximum and minimum of $\lvert\lvert \theta \lvert\lvert_2$ along an orbit are plotted, resulting in two different curves representing one solution. Multiple solutions related by symmetry, in particular those resulting from pitchfork bifurcations, share the same value of $\lvert\lvert \theta \lvert\lvert_2$. In order to distinguish between symmetry-related flow fields, we use a local measurement $\theta_{\rm local}$ based on the temperature at a single point. Here and in \citet{Zheng2023part2}, the bifurcation diagrams contain apparent intersections of curves indicating solution branches which are not related to bifurcations but result from projecting the high-dimensional flow fields onto a one-dimensional scalar quantity. Apparent intersections which are not labelled as bifurcations are of this spurious type. 

\par In addition, we also calculate the thermal energy input ($I$) due to buoyancy forces and the dissipation ($D$) due to viscosity, both averaged over the domain, for phase portrait visualizations. We refer readers to \citet{Reetz2020a} for more details. In order to visualize instantaneous flow fields or eigenvectors, we plot their temperature fields $\theta$ on the $y$-$z$ plane on the midplane at $x=0$ and/or on the $x$-$z$ plane at $y=0.5$.

\section{Equilibria}
\label{part1_Equilibria}
\par
\begin{figure}
    \centering
    \includegraphics[width=\columnwidth]{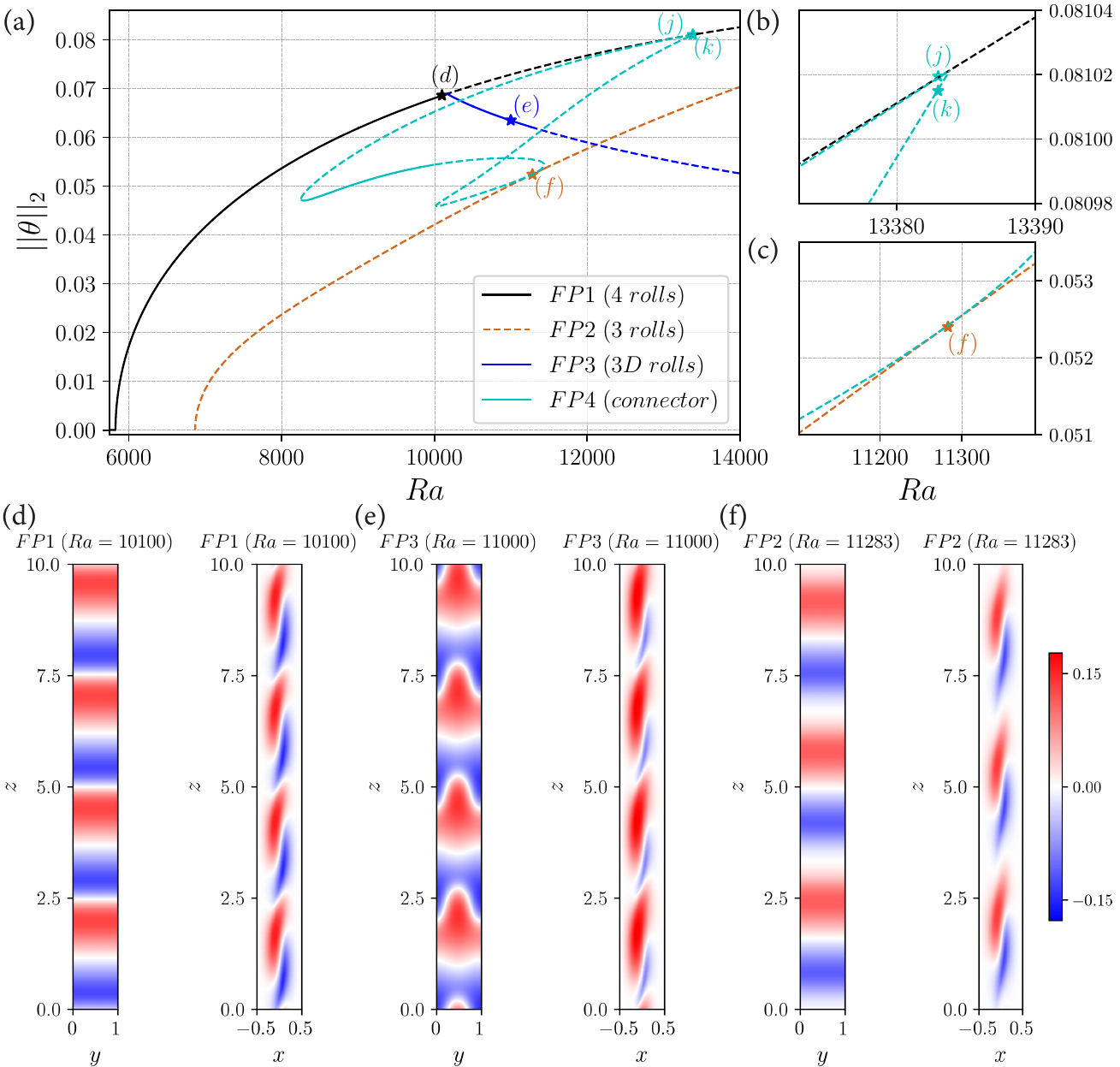}
    \captionsetup{font={footnotesize}}
    \captionsetup{width=13.5cm}
    \captionsetup{format=plain, justification=justified}
    \caption{\label{small_FP} (a) Bifurcation diagram of fixed points (FPs) using global quantity $||\theta||_2$. Solid and dashed curves signify stable and unstable states respectively. (b) and (c) zoom in on the Rayleigh number ranges within which FP4 bifurcates from FP1 and FP2. (d-f) Flow structure of equilibria visualized via the temperature field in the $y$-$z$ plane at $x=0$ and in the $x$-$z$ plane at $y=0.5$. FP1 (d), with four rolls and symmetry group $S_{FP_{1}} \equiv \braket{\pi_y, \pi_{xz}, \tau(\Delta y,2.5)}$ and FP2 (f), with three rolls and $S_{FP_{2}} \equiv \braket{\pi_y,\pi_{xz}, \tau(\Delta y,10/3)}$, both bifurcate from the conductive base flow (stable for FP1 and unstable for FP2), breaking $z$-translation symmetry. FP3 (e), with $S_{FP_{3}} \equiv \braket{\pi_y,\pi_{xz}\tau(0.5,0), \tau(0, 2.5)}$, bifurcates from FP1 and breaks its $y$-translation symmetry. FP4 (see figure \ref{small_FP_xz}), with $S_{FP_4} \equiv \braket{\pi_y, \pi_{xz}, \tau(\Delta y, 0)}$ bifurcates from FP1 at $Ra= 13383.9$ and intersects FP2 at $Ra= 11283$. The stars in (a-c) indicate where (d), (e) and (f) in the current figure as well as (f), (j) and (k) in figure \ref{small_FP_xz} are visualized.}
\end{figure}

 Our goal is to understand the formation and instabilities of convection rolls in the computational domain $[L_x, L_y, L_z] = [1, 1, 10]$, the domain studied by \citet{Gao2013}. Figure \ref{small_FP} displays the equilibria that we have studied. Many more unstable branches undoubtedly exist that are not shown in this figure, since a new branch is formed whenever the real part of an eigenvalue traverses zero. Since some of the states we discuss can also exist in domains $[1, 1, 2.5]$ and $[1, 0, 10]$, we will also mention their existence and stability ranges in these smaller domains. 

\subsection{Two primary and one secondary circle pitchfork bifurcation}
\label{primary_bif}
\par The conductive base flow becomes linearly unstable at $Ra=5826$, close to the threshold $Ra=5800$ reported by \citet{Gao2013}, where it bifurcates to a two-dimensional state containing four co-rotating transverse convection rolls. Each roll has height (or wavelength) $\Delta z=L_z/4=10/4=2.5$ and we will use both decimal and fractional notation as seems appropriate. The critical wavelength and Rayleigh number for $Pr=0.71$ computed by \cite{vest1969stability} are $2.37$ and $5595$, respectively; since our wavelength is constrained by our imposed vertical periodicity to be a divisor of 10, the threshold in $Ra$ is necessarily higher.

\par The four-roll state, called FP1 in figure \ref{small_FP}(a), is illustrated in figure \ref{small_FP}(d) and figure \ref{VC_figure}(c), which show the temperature field $\theta$. We recall that we have defined $\theta$ to be the deviation from the conductive solution, which we show in figure \ref{VC_figure}(b); the full temperature field is shown in figure \ref{VC_figure}(d). Examination of figure \ref{VC_figure} along with the corresponding velocity fields shows that the motion of the deviation fields is clockwise (figure \ref{VC_figure}(c)), but that when added to the base flow (figure \ref{VC_figure}(b)), the full motion in each roll (figure \ref{VC_figure}(d)) is counter-clockwise: colder fluid on the left ($x=-0.5$) crosses the cavity towards the right and then rises, while warmer fluid on the right ($x=0.5$) crosses towards the left and then descends. This instability is driven by the shear in the vertical velocity, in contrast to the buoyancy-driven rolls that occur in Rayleigh-B\'enard convection. FP1 has reflection and translation symmetries $S_{FP_{1}} \equiv \braket{\pi_y,\pi_{xz}, \tau(\Delta y,2.5)}\simeq [O(2)]_y \times [D_4]_{xz}$, where the translation symmetry in $L_z/4=2.5$ results from its four vertically stacked identical rolls in figure \ref{small_FP}(d).

\par We have found another fixed point, FP2, containing three identical rolls, which is shown in figure \ref{small_FP}(f). FP2 bifurcates from the unstable conductive base flow at $Ra=6868.7$ and remains unstable over its entire range of existence. FP2 is invariant under reflection and translation symmetries $S_{FP_{2}} \equiv \braket{\pi_y,\pi_{xz}, \tau(\Delta y,10/3)}\simeq [O(2)]_y \times [D_3]_{xz}$. FP1 is stable until $Ra=10166$, when it bifurcates to a state containing four three-dimensional steady rolls, which we have called FP3. This state, observed by \cite{Gao2013} and shown in figure \ref{small_FP}(e), is in turn stable until $Ra=11261$. FP3 is invariant under $S_{FP_{3}} \equiv \braket{\pi_y, \pi_{xz}\tau(0.5,0), \tau(0,2.5)} \simeq [D_1]_y \times [D_4]_{xz}$; symmetry $\tau(\Delta y,0)$ is broken at the circle pitchfork bifurcation point at $Ra = 10166$.

\subsection{FP4: connector between FP1 and FP2 states}
\label{simultan_bif_small}
\begin{figure}
    \centering
    \includegraphics[width=\columnwidth]{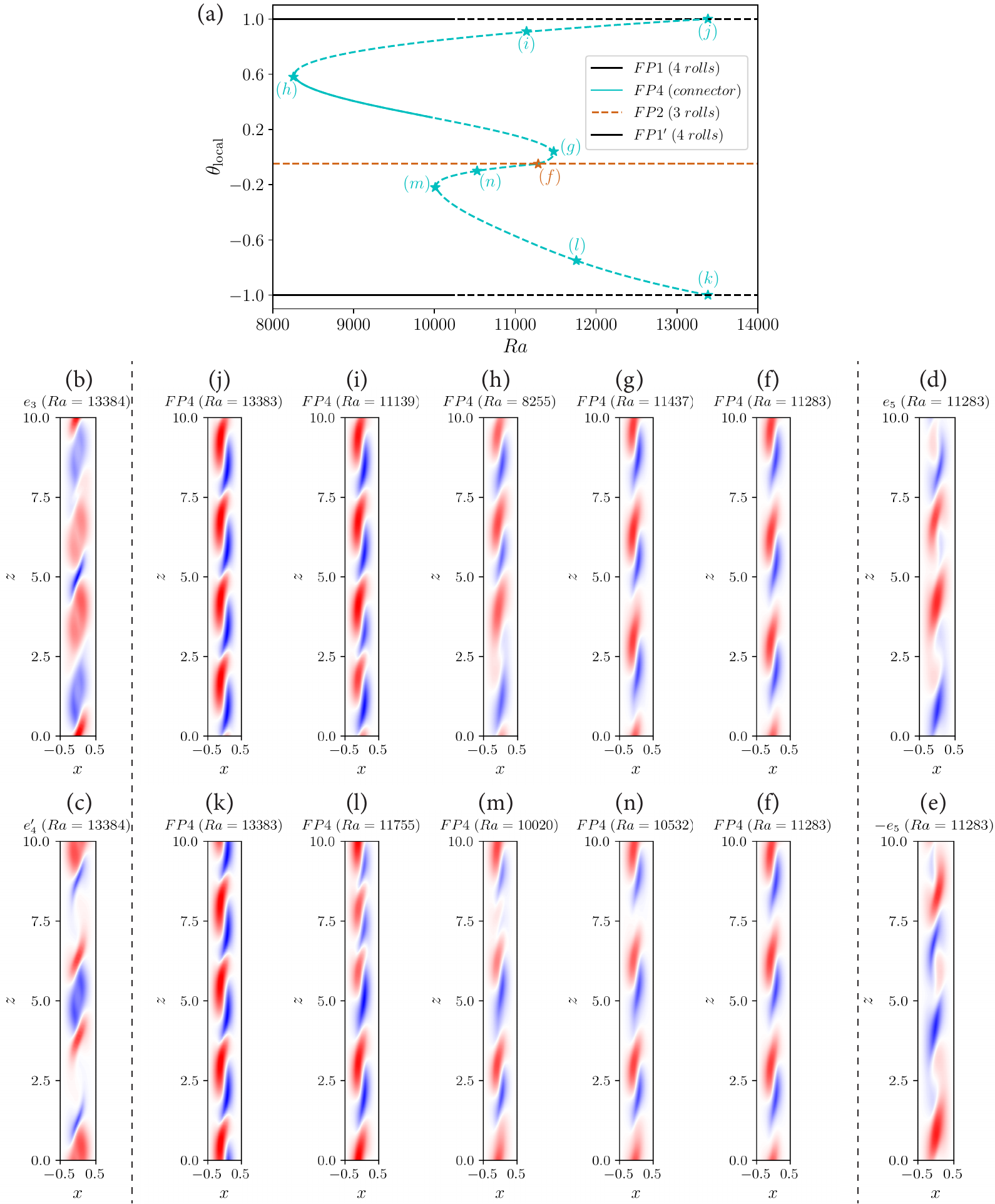}
    \captionsetup{font={footnotesize}}
    \captionsetup{width=13.5cm}
    \captionsetup{format=plain, justification=justified}
    \caption{\label{small_FP_xz} (a) Partial bifurcation diagram focusing on connector state using normalized local quantity $\theta_{\rm local}$ defined in \eqref{normalizd_thetalocal}. Solid and dashed curves signify stable and unstable states respectively. (b-e) Eigenmodes (b-c, left of dashed line and d-e, right of dashed line) and (f-n) equilibria visualized on the $x$-$z$ plane. The two ends, (j) and (k), of the connector branch FP4 are created at subcritical pitchfork bifurcations from four-roll branches FP1 and FP1$^\prime$, associated with eigenmodes (b) $e_3$ and (c) $e_4^\prime$, respectively. From (j) to (f), the rolls above and below $z=2.5$ merge, while from (k) to (f), the roll at $z=7.5$ disappears; we call these the roll-merging and roll-disappearing half-branches of FP4, respectively. At (f), the two half-branches meet three-roll branch FP2 in a transcritical bifurcation; eigenmodes (d) $e_5$ and (e) $-e_5$ lead to the roll-splitting and roll-creation portions of FP4, respectively. Solutions FP1 and FP2 have symmetry groups $[D_4]_{xz}$ and $[D_3]_{xz}$, respectively. The eigenmodes and the FP4 solutions all have the smaller symmetry group $[Z_2]_{xz}$ with no $z$-translation symmetry. (All have $[O_2]_y$.) Labels (f), (j) and (k) correspond to those used in the bifurcation diagrams in figures \ref{small_FP}(a-c). In (f-n), the same color bar is used as in figures \ref{small_FP}(d-f).}
\end{figure}

\par Figure \ref{small_FP}(a-c) shows another equilibrium, which we have called FP4, bifurcating from FP1 at $Ra= 13383.9$ and intersecting FP2 at $Ra=11283$. Two sets of solutions, labelled as (j) and (k), appear from the same FP1 state via simultaneous subcritical pitchfork bifurcations. We will call these {\it half-branches}; the reason for this and for their simultaneous bifurcation will become clear below.

\par We will need to consider another translation-symmetry related version of FP1, shifted by a half-roll ($\Delta z=\pm 1.25$) from FP1, which we will call FP1$^\prime \equiv \tau(0, 1.25)$FP1. Because the global quantity $||\theta||_2$ cannot distinguish between symmetry-related states, we represent FP4 in figure \ref{small_FP_xz}(a) by the local and normalized quantity
\begin{align}
    \theta_{\rm local}(Ra)\equiv\left.\frac{\theta(Ra)}{|\theta(Ra=13383)|}\right\vert_{x=0,z=4.375} \in [-1,1].
    \label{normalizd_thetalocal}
\end{align}
To emphasize the variation of $\theta_{\rm local}$ as FP4 is traversed, for visualization, we suppress the variation along the FP1 and FP2 branches.

\par The two endpoints of the FP4 branch, related by a half-roll shift of $\Delta z = 1.25$, are shown in figures \ref{small_FP_xz}(j) and (k). In the bifurcations from FP1 to FP4, the four-fold translational symmetry in $z$ is lost, but $(x,z)$ reflection symmetry is retained, leading to $S_{FP_4} \equiv \braket{\pi_y, \pi_{xz}, \tau(\Delta y, 0)} \simeq [O(2)]_y \times [Z_2]_{xz}$. We have chosen the spatial phase such that the centers of symmetry of figures \ref{small_FP_xz}(j) and (k) are located at $z$ values that are multiples of $10/8=1.25$. During the numerical continuation of the FP4 branch, the phase in $z$ has been fixed by imposing two reflection symmetries.

\subsubsection{Tour of FP4: two methods for eliminating one roll} 
\label{tour}
\par We begin our tour of FP4 from figure \ref{small_FP_xz}(j), which displays one end of the FP4 branch or, equivalently, FP1. Going from figure \ref{small_FP_xz}(j) to \ref{small_FP_xz}(i), the between-roll boundary at $z=2.5$ becomes weaker. In contrast, at $z=7.5$, the roll boundary is strengthened while the far edges of the two surrounding rolls are weakened. By figure \ref{small_FP_xz}(h), the two rolls formerly surrounding $z=2.5$ have merged into a single large roll. For this reason, we call (j)-(i)-(h)-(g)-(f) in figure \ref{small_FP_xz}(a) the \emph{roll-merging half-branch}. Starting from FP1$^\prime$ in figure \ref{small_FP_xz}(k), the opposite endpoint of the FP4 branch, the roll centered around $z=7.5$ weakens in figure \ref{small_FP_xz}(l) and has almost disappeared by the saddle-node bifurcation of figure \ref{small_FP_xz}(m). We call (k)-(l)-(m)-(n)-(f) in figure \ref{small_FP_xz}(a), the \emph{roll-disappearing half-branch}.

\par At $Ra=11283$, figure \ref{small_FP_xz}(f) has three equally spaced rolls and belongs to branch FP2. This is why we choose to call this state the dividing point of branch FP4 into two half-branches. The meeting between FP2 and the two half-branches is a transcritical bifurcation which will be the topic of \S \ref{transcritical_section}. Both FP4 half-branches lead from four rolls to three rolls, but in different ways. In the pathway from figures \ref{small_FP_xz}(j) to (f), the space between two rolls blurs and the two rolls merge. In the pathway from figures \ref{small_FP_xz}(k) to (f), one roll weakens and disappears. These two types of transitions can occur at any of the four roll centers and roll boundaries. Thus eight half-branches bifurcate simultaneously from any FP1 state: four roll-merging half-branches like figures \ref{small_FP_xz}(j) to (f) and four roll-disappearing half-branches like figures \ref{small_FP_xz}(k) to (f). These eight branches connect an FP1 state with its half-roll-shifted state FP1$^\prime$.

\par
\begin{figure}
    \centering
    \includegraphics[width=0.33\columnwidth]{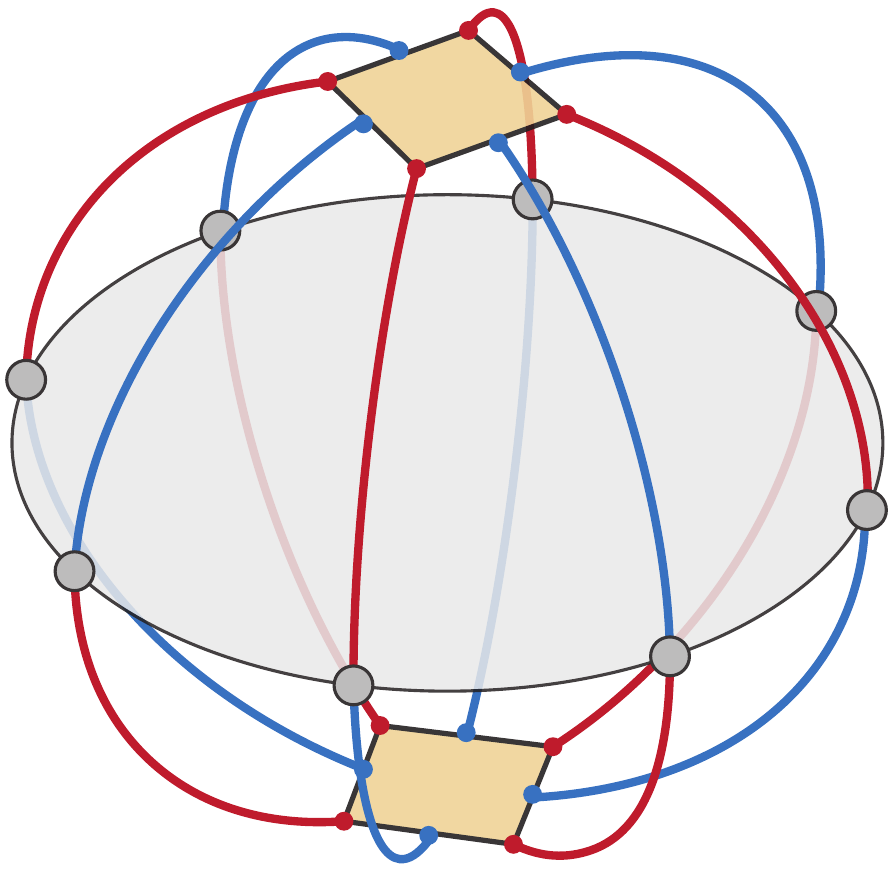}
    \captionsetup{font={footnotesize}}
    \captionsetup{width=13.5cm}
    \captionsetup{format=plain, justification=justified}
    \caption{\label{D4_schematic} Schematic diagram of the set of FP4 branches associated with figure \ref{small_FP_xz}. The square on the top represents the pitchfork bifurcation point of FP1, figure \ref{small_FP_xz}(j), while the square on the bottom, rotated by $2\pi/8$ with respect to the top one, represents that of FP1$^\prime$, figure \ref{small_FP_xz}(k). Four roll-merging half-branches, shown in red, emanate from four corners of each of the squares; and four roll-disappearing half-branches, shown in blue, emanate from four sides of each of the squares. These are the half-branches shown in figures \ref{small_FP_xz}(j)-(i)-(h)-(g)-(f) and \ref{small_FP_xz}(k)-(l)-(m)-(n)-(f), and also those obtained by $\tau(0,2.5)$, $\tau(0,5.0)$, and $\tau(0,7.5)$, in which the roll merging or disappearing occurs at other locations. Each roll-merging half-branch emanating from FP1 meets a roll-disappearing half-branch emanating from FP1$^\prime$, and vice versa, at the equator, on which are situated the transcritical bifurcation points with the FP2 branch, such as figure \ref{small_FP_xz}(f).}
\end{figure}

 This scenario is schematized in figure \ref{D4_schematic}. Each line of longitude (meridian) on the globe-like figure represents a branch connecting FP1 (top square) and FP1$^\prime$ (bottom square), like that shown in figure \ref{small_FP_xz}. The roll-merging half-branches are colored in red and emerge from the corners of a square, the roll-disappearing half-branches in blue and emerge from the sides of a square. The fact that four of each emerge at each of the squares corresponds to the fact that each of FP1 and FP1$^\prime$ contains four rolls and four inter-roll spaces which can undergo roll-disappearance or roll-merging. Each half-branch of one color emanating from FP1 meets a half-branch of the opposite color emanating from FP1$^\prime$ at the equator, which contains transcritical bifurcation points of different phases in $z$. 

\subsubsection{Eigenvectors of FP1 and FP1$^\prime$}
\par Figures \ref{small_FP_xz}(b) and (c) show the unstable eigenmodes $e_3$ of FP1 and $e_4^\prime$ of FP1$^\prime$ at $Ra=13384$ that are responsible for the two simultaneous subcritical pitchfork bifurcations that create the two half-branches of FP4. We call these $e_3$ and $e_4^\prime$ because the FP1 (and FP1$^\prime$) branch at $Ra=13384$ has two larger positive eigenvalues resulting from the circle pitchfork bifurcation to FP3. $e_3$ and $e_4^\prime$ have the same eigenvalue, $\lambda_{3,4}$. The Arnoldi method computes two linearly independent eigenmodes of the double eigenvalue $\lambda_{3,4}$; we select the linear combination of these that most resembles the difference between FP4 and FP1 at the bifurcation point of figure \ref{small_FP_xz}(j). The eigenvectors of FP1$^\prime$ are related to those of FP1 by a translation $\tau(0,1.25)$. We select as $e_4^\prime$ the analogous combination of eigenmodes of FP1$^\prime$ that most resembles the difference between FP4 and FP1$^\prime$ at the bifurcation point of figure \ref{small_FP_xz}(k). Eigenmodes $e_3$ and $e_4^\prime$ differ qualitatively: $e_3$ has two narrow intense temperature extrema surrounded by wide diffuse patches of the opposite sign, while $e_4^\prime$ has two wide diffuse patches surrounded by narrow extrema. Each eigenmode has two centers of $\pi_{xz}$ symmetry, at $z=2.5$ and $7.5$.

\par Examining the eigenvectors helps to understand the progression from the four-fold translation-symmetric FP1 (and FP1$^\prime$) to FP4. The eigenvectors describe defects that, when added to FP1 (or FP1$^\prime$), lead to roll merging or roll disappearance. The red (hot) and blue (cold) diffuse patches of the $e_3$ are in opposition to those of FP1 at the boundary between two rolls at $z=2.5$; compare figures \ref{small_FP_xz}(b) and \ref{small_FP_xz}(j). This implies that the between-roll boundary at $z=2.5$ becomes weaker along this half-branch. In contrast, at $z=7.5$, FP1 and $e_3$ have temperatures of the same sign, so this roll boundary is strengthened. Turning now to the pathway (k)-(l)-(m)-(n)-(f), this roll-disappearing half-branch is associated with eigenvector $e_4^\prime$ in figure \ref{small_FP_xz}(c). $e_4^\prime$ is very weak at $z=2.5$ and at $z=7.5$, around which rolls of FP1$^\prime$ are centered. However, the temperature of $e_4^\prime$ surrounding $z=2.5$ is such as to reinforce the roll at $z=2.5$ of FP1$^\prime$, whereas $e_4^\prime$ and FP1$^\prime$ display opposite temperatures surrounding $z=7.5$. The roll of FP1$^\prime$ at $z=7.5$ consequently disappears along this half-branch.

\par Eigenmodes of the Jacobian matrix describe the temporal dynamics near a fixed point $\bar{u}$, but we have used them above to describe the tangent along a branch (or half-branch) near a bifurcation. We now explain the justification for this. For a dynamical system $du/dt=f(u,Ra)$, a curve of fixed points $\bar{u}(Ra)$ is defined via $0=f(\bar{u}(Ra),Ra)$. Differentiating in $Ra$ yields
\begin{align}
    0&=\frac{\partial f}{\partial Ra} + [Df]_{\bar{u}}\frac{d\bar{u}}{dRa},
\end{align}
where $[Df]_{\bar{u}}$ is the Jacobian evaluated at $\bar{u}$. Near a bifurcation, the Jacobian has an eigenvalue $\lambda_{\rm bif}$ near zero so that multiplication by the inverse Jacobian projects onto the bifurcating eigenvector $e_{\rm bif}$:
\begin{align}
    \frac{d\bar{u}}{dRa}&=-[Df]_{\bar{u}}^{-1}\frac{\partial f}{\partial Ra} 
    = -\sum_j \frac{1}{\lambda_j}\left\langle\frac{\partial f}{\partial Ra}, e_j\right\rangle e_j
    \approx -\frac{1}{\lambda_{\rm bif}}\left\langle\frac{\partial f}{\partial Ra},e_{\rm bif}\right\rangle e_{\rm bif},
\end{align}
where $\langle,\rangle$ is an inner product and $(\lambda_j,e_j)$ are the eigenpairs of $\left[Df\right]_{\bar{u}}$, with $|1/\lambda_{\rm bif}| \gg |1/\lambda_j|$ for the other eigenvalues of $\left[Df\right]_{\bar{u}}$. This leads to the expression
\begin{align}
\bar{u}(Ra - \Delta Ra) \approx \bar{u}(Ra) - \Delta Ra\frac{d\bar{u}}{dRa}  
\approx \bar{u}(Ra) +\frac{\Delta Ra}{\lambda_{\rm bif}}\left\langle\frac{\partial f}{\partial Ra},e_{\rm bif}\right\rangle e_{\rm bif}
\label{ebif}
\end{align}
for the evolution of a branch near a bifurcation.

\subsubsection{Normal form of $D_4$ symmetry}
\label{D4_nf}
\par The simultaneous occurrence of two pitchfork bifurcations described above is precisely the scenario seen in pattern formation on a square domain which, like FP1 (and FP1$^\prime$), has the symmetry group $D_4$, generated by $\pi_{xz}$ and $\tau(0,2.5)$. Instead of considering the FP4 branch with its endpoints FP1 and FP1$^\prime$ as we did in figure \ref{small_FP_xz}, we now consider a single phase of FP1 and its two bifurcations to roll-merging and roll-disappearing half-branches corresponding to its eigenvectors $e_3$ and $e_4$. In the square, rolls can be oriented horizontally or vertically, and these are equivalent because they are related by a rotation by $\pi/2$. The eigenvectors associated with vertical and horizontal rolls can also be combined to form diagonal eigenvectors. The nonlinear equations that are equivariant (compatible) with $D_4$ symmetry predict the existence of branches of diagonal states \citep{swift1985bifurcation, bergeon2001three} that originate from eigenvectors that are equal superpositions of vertical and horizontal eigenmodes, as will be discussed below. The diagonal roll branches bifurcate simultaneously with the horizontal and vertical roll branches, but the nonlinear diagonal states are not related to the horizontal or vertical states by symmetry operations and are therefore not equivalent. Both types of branches have a reflection symmetry, vertical or horizontal in one case, and diagonal in the other case, so that their symmetry groups are $Z_2$.

\par This scenario for pattern formation on a square domain also exists for the FP1 branch, with the four co-rotating rolls playing the role of the four sides of a square and the four inter-roll intervals playing the role of the corners. Four bifurcating branches resembling figures \ref{small_FP_xz}(j-f) result from eigenvector $e_3$ along with shifted and reflected versions, and four branches resembling figures \ref{small_FP_xz}(k-f) result from $e_4$ along with shifted and reflected versions. The bifurcations occur at the same value of Rayleigh number but the branches are not equivalent, as seen in figures \ref{small_FP}(a) and \ref{small_FP_xz}(a), for example by the different locations of the saddle-node bifurcations emanating from FP1 and from FP1$^\prime$.

\par We now give a more quantitative explanation of this scenario. Consider the dynamical system governing $(p,q)\in \mathcal{R}^2$:
\begin{subequations}
\label{D4_eqns}
\begin{align}
\dot{p} &= \left(\mu - a p^2 - b q^2\right) p,\\
\dot{q} &= \left(\mu - b p^2 - a q^2\right) q,
\end{align}
\end{subequations}
with $\mu,a,b$ all real parameters. The bifurcation parameter is $\mu$ and $a,b$ are nonlinear coefficients that saturate the instability. System \eqref{D4_eqns} is a projection of a larger system near a bifurcation onto the bifurcating eigenmodes. A normal form is the smallest system, in terms of both number of variables and polynomial order, that is able to reproduce the behavior of the larger system near the bifurcation. The form of system \eqref{D4_eqns} is dictated by the requirements that it be equivariant under (consistent with) change in sign of $p$ or $q$ and interchange of $p$ and $q$, which defines the group $D_4$. 

\begin{figure}
    \centering
    \includegraphics[width=0.85\columnwidth]{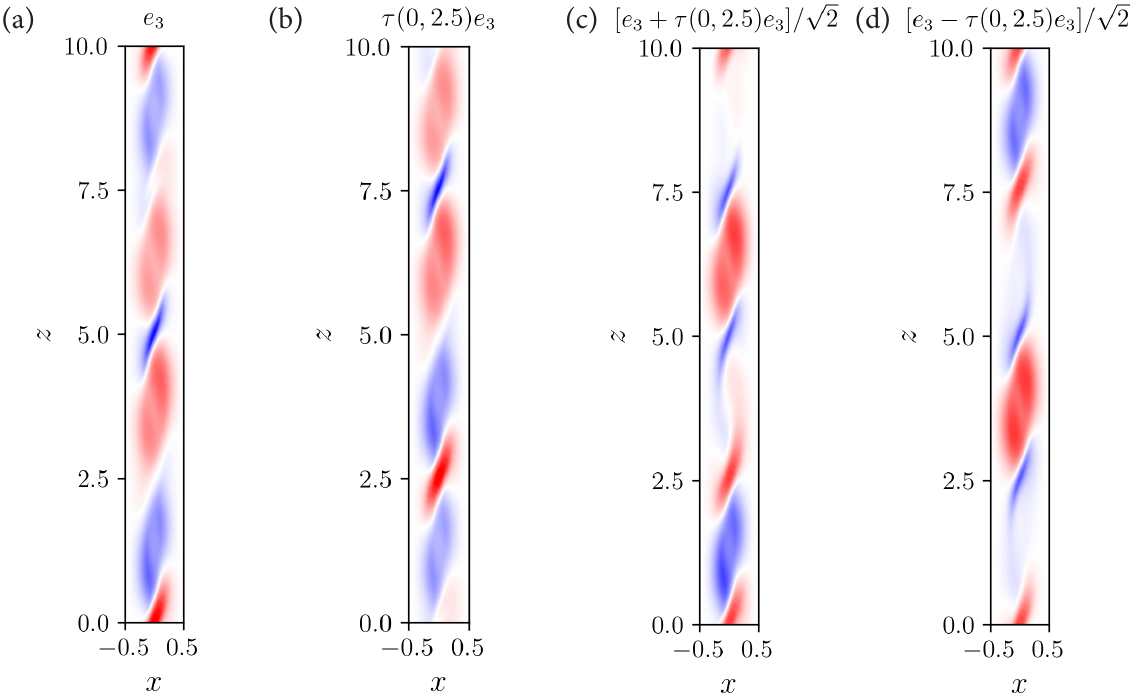}
    \captionsetup{font={footnotesize}}
    \captionsetup{width=13.5cm}
    \captionsetup{format=plain, justification=justified}
    \caption{\label{D4_small} (a) Eigenmode $e_3$ of FP1, with the same phase as in figure \ref{small_FP_xz}(b). (b) Quarter-domain-translated version of (a): $\tau(0,2.5)e_3$. (c-d) Superpositions $(e_3 \pm \tau(0,2.5)e_3)/\sqrt{2}$. Note that $(e_3+\tau(0,2.5)e_3)/\sqrt{2} = e_4= \tau(0,-3.75) e_4^\prime$ (compare with figure \ref{small_FP_xz}(c)).}
\end{figure}

\par The Jacobian of \eqref{D4_eqns} is:
\begin{align}
\left[\begin{array}{cc} 
    \mu -3ap^2 -bq^2 & -2bpq\\ 
    -2bpq & \mu -bp^2 - 3aq^2   
    \end{array}\right].
\end{align}
Evaluated at the trivial solution $(p,q)=(0,0)$, this becomes $\mu{\bf I}$, i.e. a double eigenvalue $\mu$. The non-trivial fixed points of system \eqref{D4_eqns} are:
\begin{subequations}
\label{D4_sol}
\begin{align}
  p &= \pm\sqrt{\mu/a} & q &= 0, \label{D4_sola}\\
  p &= 0 & q&=\pm\sqrt{\mu/a}, \label{D4_solb}\\
  p &= \pm\sqrt{\mu/(a+b)} & q& = \pm\sqrt{\mu/(a+b)}, \label{D4_solc}\\
  p &= \pm\sqrt{\mu/(a+b)} & q& = \mp\sqrt{\mu/(a+b)}. \label{D4_sold}
\end{align}
\end{subequations}
Thus \eqref{D4_eqns} has eight non-trivial solutions, two each of types \eqref{D4_sola}, \eqref{D4_solb}, \eqref{D4_solc} and \eqref{D4_sold}. Although solutions \eqref{D4_sola} and \eqref{D4_solb} are related to one another by the symmetry $(p,q) \rightarrow (-q,p)$, as are solutions \eqref{D4_solc} and \eqref{D4_sold}, solutions \eqref{D4_solc} and \eqref{D4_sold} are not related to solutions \eqref{D4_sola} and \eqref{D4_solb} by interchanging $p$ and $q$ or by changing their signs.

\par The scenario by which FP1 gives rise to FP4 is analogous to system \eqref{D4_eqns} and \eqref{D4_sol}, with FP1 playing the role of the trivial solution $p=q=0$. The assumption of normal form theory is that FP4 solutions can be approximated as superpositions of the base state FP1 and its eigenvectors $e_3$ and $\tau(0,2.5)e_3$ at the bifurcation:
\begin{align}
{\rm FP4} = {\rm FP1} + p(t) e_3 + q(t) \tau(0,2.5)e_3,
\end{align}
with $p(t)$ and $q(t)$ governed by the amplitude equations or normal form \eqref{D4_eqns}. The quantity $p$ measures the amplitude of eigenvector $e_3$ in figure \ref{small_FP_xz}(b), which gives rise to the half-branch in which two rolls merge at $z=2.5$. The phase-shifted $\tau(0,2.5)e_3$, whose amplitude is measured by $q$, gives rise to a different half-branch in which roll-merging occurs at $z=5.0$. Figure \ref{D4_small}(a-b) shows these eigenvectors, while figure \ref{D4_small}(c) shows their normalized sum. Further shifts, $\tau(0,5)e_3 = -e_3$ and $\tau(0,7.5)e_3 = -\tau(0,2.5)e_3$ correspond to $-p$ and $-q$, respectively.

\par Turning now to the four roll-disappearing half-branches bifurcating from FP1, these are produced by eigenvectors $\tau(0, 2.5 n)e_4$ for $n=0,1,2,3$. The normalized sum of the roll-merging eigenvectors $(e_3 + \tau(0, 2.5)e_3)/\sqrt{2}$ turns out to be the roll-disappearing eigenvector $e_4$, analogously to the fact that the sum of a $p$ and a $q$ solution yields a $p+q$ solution. (Because these are eigenvectors, their amplitudes have no importance.) The sum of two neighboring roll-disappearing eigenvectors (not shown) is a roll-merging eigenvector, analogously to the combinations $(p+q)+(p-q)\propto p$ and $(p+q)-(p-q)\propto q$. This confirms the correspondence between the normal form \eqref{D4_eqns} and our hydrodynamic system with its four-roll branch FP1, its connector branch FP4 and its eigenvectors $e_3$ and $e_4$.

\subsubsection{Transcritical bifurcation between FP4 and $D_3$-symmetric FP2}
\label{transcritical_section}
\begin{figure}
    \centering
    \includegraphics[width=\columnwidth]{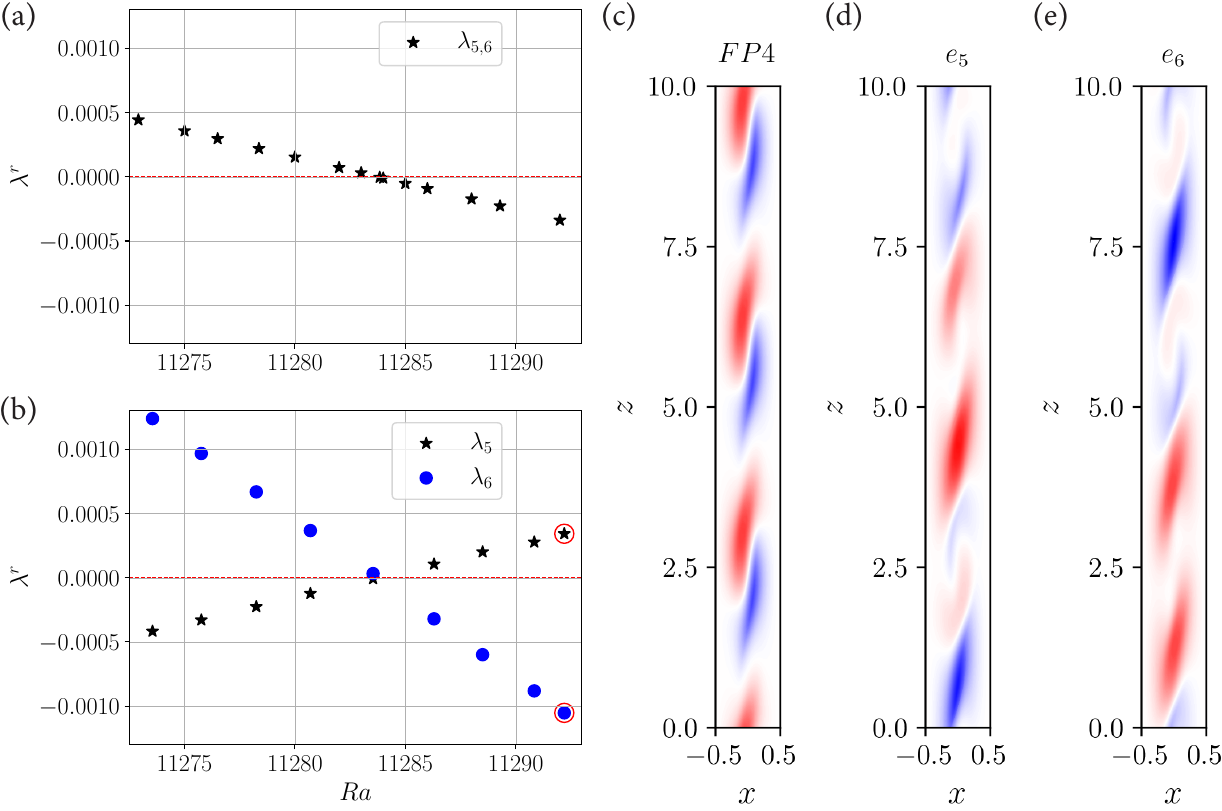}
    \captionsetup{font={footnotesize}}
    \captionsetup{width=13.5cm}
    \captionsetup{format=plain, justification=justified}
    \caption{\label{FP2-FP4-eigen} (a-b) Evolution of eigenvalues relevant to transcritical bifurcation FP2$\leftrightarrow$FP4 at $Ra=11283$. (a) bifurcating double eigenvalue $\lambda_{5,6}$ of FP2. (b) Two eigenvalues of FP4: $\lambda_5$ (whose sign is reversed with respect to $\lambda_{5,6}$ of FP2) and $\lambda_6$, where $|\lambda_6| \approx 3|\lambda_5|$. (c) Base state FP4 for eigenmodes $e_5$ and $e_6$ in (d-e). (d-e) Eigenmodes $e_5$ and $e_6$ of FP4 associated with $\lambda_5$ and $\lambda_6$ at $Ra=11292.2$ (red circles in (b)). Eigenmode $e_5$ is $xz$-reflection symmetric about $z=2.5$ (and $z=7.5$) and is related to a change in amplitude along the branch while $e_6$ is anti-$xz$-reflection symmetric about $z=2.5$ (and $z=7.5$) and is related to a change in phase perpendicular to the branch. In (c), the same color bar is used as in figures \ref{small_FP} and \ref{small_FP_xz}.}
\end{figure}

\par Figure \ref{small_FP_xz}(a) shows the intersection between FP4 and the three-roll branch FP2 at figure \ref{small_FP_xz}(f) in a transcritical bifurcation (TC). From the point of view of FP2, the FP4 roll-merging half-branch (j)-(i)-(h)-(g)-(f) can be called a roll-splitting half-branch when traversed in the opposite order (f)-(g)-(h)-(i)-(j). Similarly, the FP4 roll-disappearing half-branch (k)-(l)-(m)-(n)-(f) can be called a roll-creation half-branch when traversed in the order (f)-(n)-(m)-(l)-(k). Because FP2 has three-fold translation symmetry $\tau(0,10/3)$, any of the three rolls in figure \ref{small_FP_xz}(f) can be the site of a roll-splitting or a roll-creation event, and so six FP4 half-branches, three of each type, emanate from FP2 at TC. These join pairwise at FP2: for example, in figure \ref{small_FP_xz}, the upper half-branch with roll-splitting at $z=2.5$ (figure \ref{small_FP_xz}(h) and (i)) meets the lower half-branch in which roll-creation occurs at $z=7.5$ (figure \ref{small_FP_xz}(m) and (l)). (We assume that the saddle-node bifurcations have no effect on this scenario.)

\par The bifurcation from FP2 to FP4 breaks three-fold translation symmetry but retains reflection symmetry $\pi_{xz}$. This can be seen in figure \ref{small_FP_xz}(h), for example, where the roll centered at $z=2.5$ is stretched, while the other two rolls remain of the same size and related to one another by reflection in $z=2.5$. For figure \ref{small_FP_xz}(m), the reasoning is the same, but is applied to the inter-roll space at $z=7.5$.

\par We now turn to the eigenmodes of FP2 and FP4 close to TC. To the right of figure \ref{small_FP_xz}(f), figure \ref{small_FP_xz}(d) displays the eigenmode $e_5$ of FP2 at $Ra\gtrsim11283$ leading to FP4. As previously, the name $e_5$ is used because FP2 has four eigenvalues with larger real parts. By a slight abuse of notation, we use $-e_5$ to denote the direction in which FP2 is approached from FP4 for $Ra\lesssim11283$ and visualize it in figure \ref{small_FP_xz}(e). We obtain $\pm e_5$ by subtracting the temperature fields of FP2 from FP4 states to the right and left of point (f) in figure \ref{small_FP_xz}(a), as well as from the Arnoldi method. Using \eqref{ebif} again, we can superpose FP2 at figure \ref{small_FP_xz}(f) with eigenvector $e_5$ at figure \ref{small_FP_xz}(d) to yield FP4 at figures \ref{small_FP_xz}(g-h), since $e_5$ opposes the roll in FP4 centered at $z=2.5$, leading to an expanded roll and eventually to roll-splitting. Similarly, we can superpose figure \ref{small_FP_xz}(f) with eigenvector $-e_5$ at figure \ref{small_FP_xz}(e) to yield FP4 at figures \ref{small_FP_xz}(n-m). Since $-e_5$ opposes the rolls on either side of $z=7.5$ in FP4, this inter-roll space expands, eventually making room for roll-creation.

\par As FP2 has three-fold translation symmetry, $e_5$ of FP2 can be shifted by $\pm 10/3$ in $z$, yielding a triple of eigenvectors only two of which are linearly independent, since $\tau(0,10/3)e_5 + \tau(0,-10/3)e_5 = -e_5$. These share the same eigenvalue $\lambda_{5,6}$, depicted in figure \ref{FP2-FP4-eigen}(a). Along branch FP4, these eigenvectors are modified, so that they are no longer related by $\tau(0,\pm 10/3)$ and hence have different eigenvalues, shown as $\lambda_5$ and $\lambda_6$ in figure \ref{FP2-FP4-eigen}(b). Eigenvectors $e_5$ and $e_6$ of FP4 at $Ra=11292.2$, shown in figures \ref{FP2-FP4-eigen}(d) and (e) are symmetric and anti-symmetric, respectively, with respect to $xz$-reflection symmetry about $z=2.5$ and $z=7.5$. 

\subsubsection{$D_3$ symmetry}
\label{D3_nf}
\par
\begin{figure}
    \centering
    \includegraphics[width=\columnwidth]{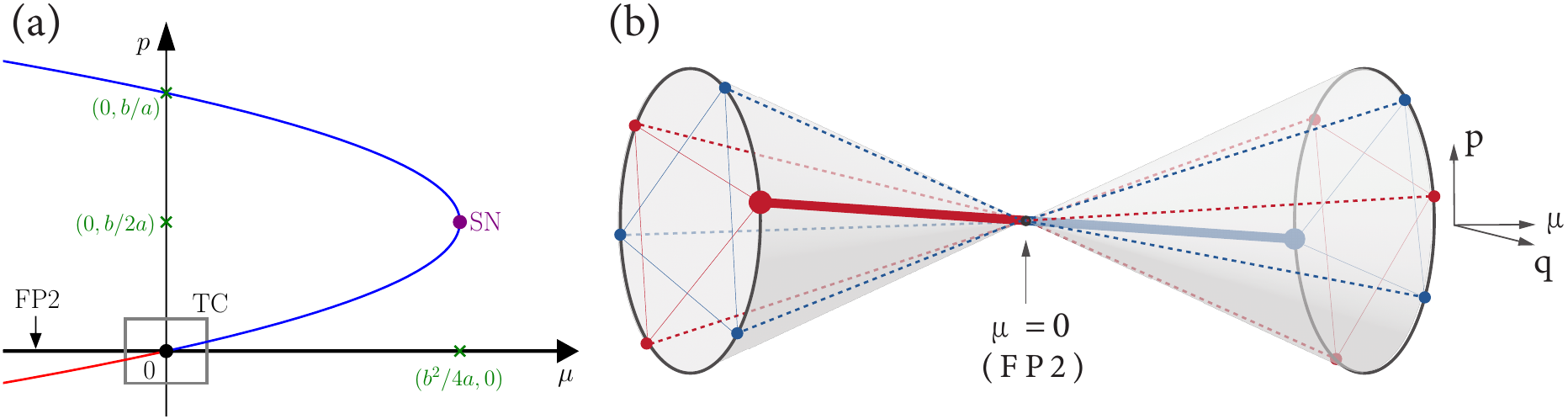}
    \captionsetup{font={footnotesize}}
    \captionsetup{width=13.5cm}
    \captionsetup{format=plain, justification=justified}
    \caption{\label{D3_schematic} (a) Parabola in the $(\mu,p)$ plane representing the FP4-like solutions of $\mu - bp + ap^2=0$ (with $a,b$ positive) of \eqref{D3_normalform}. The thicker horizontal axis ($p=0$) represents the FP2 branch. The saddle-node SN at $\mu=b^2/(4a)$ and transcritical bifurcation TC at $\mu=0$ are marked by circles. The region surrounded by the grey square corresponds to the meeting point of two types of half-branches (red and blue) and to the double-cone schematic in (b). (b) Three-dimensional schematic figure in the $(p,q,\mu)$ plane illustrating the solutions near TC. Each cone contains three FP4 solutions of each type (red or blue), with angular (phase) separation $2\pi/3$. These all intersect the $\mu$ axis representing FP2 at $\mu=0$ in a transcritical bifurcation. The solid thicker line (red on the left cone and blue on the right) represents the path of figure \ref{small_FP_xz}, while the five dashed lines correspond to paths that would be observed with a $\Delta z=\pm 10/3$ shift in $z$ or a reversal of the path direction or both.}
\end{figure}

 We now consider the equations governing bifurcation in the presence of $D_3$ symmetry. In this case, we will not consider the normal form, but a related system, i.e.\ the universal unfolding of the degenerate case of the normal form, because these are the equations which best describe our results. See \citet{gambaudo1985perturbation, golubitsky1988singularities, Dawes2005} for details. These equations are:
\begin{subequations}
\label{D3_normalform}
\begin{align}
\dot{p} &= -\mu p + b(p^2-q^2) - ap(p^2+q^2), \\
\dot{q} &= -\mu q - 2bpq - aq(p^2+q^2),
\end{align}
\end{subequations}
with $a, b$ real parameters, $\mu$ the bifurcation parameter, and $(p,q)$ the amplitudes of eigenmodes of the FP2 branch. As stated in \S \ref{D4_nf}, the correspondence of \eqref{D3_normalform} with our high-dimensional hydrodynamic system consists of approximating FP4 by a superposition of FP2 with its eigenvectors $e_5$ and $\tau(0, 10/3)e_5$ at the bifurcation point, whose amplitudes are represented here by $p$ and $q$:
\begin{align}
{\rm FP4} = {\rm FP2} + p(t) e_5 + q(t) \tau(0, 10/3)e_5.
\label{D3_near_bif}
\end{align}
Let us begin by studying steady solutions with $q=0$:
\begin{subequations}
\begin{align} 
p&= 0, \\
\mu - bp + ap^2 = 0  \Rightarrow 
p &=\left[b \pm \sqrt{b^2 - 4\mu a}\right]/(2a). 
\label{d3sol}
\end{align}
\end{subequations}
These are shown in figure \ref{D3_schematic}(a). The trivial solution $p=0$ corresponds to the FP2 branch. Two sets of solutions corresponding to FP4 are created at $\mu=b^2/(4a)$; this is a saddle-node bifurcation. The set of solutions closer to zero intersects the trivial FP2-type branch $p=0$ in a transcritical bifurcation at $\mu=0$. These two bifurcations are marked as SN and TC in the parabola in figure \ref{D3_schematic}(a). The saddle-node and transcritical bifurcations are also seen in the hydrodynamic case and are labelled by (g) and (f) in the bifurcation diagram of figure \ref{small_FP_xz}.

\par The Jacobian of \eqref{D3_normalform} is:
\begin{align}
\left[\begin{array}{cc} 
    -\mu + 2bp - 3ap^2 - aq^2 & -2bq -2apq\\ 
    -2bq -2apq & -\mu - 2bp - 3aq^2 - ap^2   
    \end{array}\right].
\end{align}
Evaluated at $(p,q)=(0,0)$ corresponding to FP2, this becomes $-\mu{\bf I}$, i.e. one double eigenvalue $-\mu$. Evaluated at the $q=0$, $\mu - bp+ ap^2=0$ solution corresponding to FP4, we obtain
\begin{align}
    \left[\begin{array}{cc} 
    -\mu + 2bp - 3(bp-\mu) & 0 \\
    0 & -\mu - 2bp +(\mu-bp)
    \end{array}\right] =
    \left[\begin{array}{cc} 
    2\mu - bp & 0 \\
    0 & -3bp
    \end{array}\right].
\end{align}
Since $ap^2$ can be neglected near TC, the FP4-type solutions \eqref{d3sol} take the form $p\approx \mu/b$. (In accordance with the previous nomenclature, these are two half-branches, one for $\mu>0$ and the other for $\mu<0$.) The Jacobian becomes
\begin{align}
    \left[\begin{array}{cc} 
    2\mu - \mu & 0 \\ 0 & -3\mu
    \end{array}\right] =
    \left[\begin{array}{rc} 
    \mu & 0 \\ 0 & -3\mu 
    \end{array}\right].
\end{align}
Thus, the FP4 states emanating from TC each have two eigenvalues of opposite signs, which are approximately $\mu$ (in the $p$-direction connecting FP2 and FP4) and $-3\mu$ (in the $q$-direction, perpendicular to $p$). This is precisely the behavior seen in figure \ref{FP2-FP4-eigen}(a)-(b). Indeed, if we define $-\mu$ to be the eigenvalue of FP2 in figure \ref{FP2-FP4-eigen}(a), then we find that the eigenvalues of FP4 in figure \ref{FP2-FP4-eigen}(b) are approximately $\mu$ and $-3\mu$.  

\par Dropping now the requirement that $q=0$, two more solutions to \eqref{D3_normalform} of FP4-type exist, related to the $q=0$ solution by rotation by $\pm 2\pi/3$ in the $(p,q)$ plane:
\begin{align}
\left(\begin{array}{c} p \\ q \end{array}\right) 
\rightarrow \left(\begin{array}{rr} 
\cos(2\pi/3) & \pm \sin(2\pi/3) \\
\mp \sin(2\pi/3) & \cos(2\pi/3) 
\end{array}\right)
\left(\begin{array}{c} p\\q \end{array}\right).
\label{D3_other2sols}
\end{align}
Each can be assigned an amplitude $\sqrt{p^2+q^2} \approx \mu/b$ and a phase $[\tan^{-1}(q/p)]$. The phase can in turn be mapped to a vertical location in $[0,L_z)$ of a defect in one of the three rolls or inter-roll spaces. Such a defect is a precursor to roll-splitting or roll-creation, respectively, as one leaves TC along one of the half-branches. The eigenvalues of the other two FP4 solutions are again $\mu$ and $-3\mu$. The eigenvector associated with $\mu$ resembles the defect itself; i.e.\ it corresponds to a change in its amplitude. The eigenvector associated with $-3\mu$ corresponds to a change in phase, i.e.\ a tendency for the defect to translate in $z$. Like all eigenvectors, this tendency is local to TC and nonlinear trajectories deviate from the phase-translation path before any phase change is actually achieved.

\par A schematic illustration of these solutions and the transcritical bifurcation is shown in figure \ref{D3_schematic}(b). Roll-splitting and roll-creation half-branches are shown in red and blue respectively. Three of each type of half-branch exist on each cone. The thick red and blue half-branches comprise the branch followed in figure \ref{small_FP_xz} from FP1 to FP1$^\prime$, along which roll-merging and then roll-creation occur. Another two half-branches comprise a branch from FP1 to FP1$^\prime$ along which roll-disappearance and then roll-splitting occur. The other four branches have starting or ending points that are shifted by $\Delta z\pm 10/3$ from FP1 or FP1$^\prime$. These six branches are all included in figure \ref{D3_schematic}(b) in order to give a full picture of the transcritical bifurcation; their depiction is not necessary for the understanding of the single path of figure \ref{small_FP_xz}. 

\subsection{Wider perspectives}
\par In the previous subsections, we have extensively discussed the $D_4$ and $D_3$ bifurcation scenarios separately. We now take a wider perspective, discussing various aspects of the interaction between the $D_4$ and $D_3$ bifurcations.

\par The double-cone visualization of figure \ref{D3_schematic} may seem incompatible with the globe-like visualization of figure \ref{D4_schematic}. In fact, each figure is local and the two visualizations have only two branches in common. Each branch belongs simultaneously to a sphere and to a double-cone. Four of the branches traversing TC through the double-cone are not present in figure \ref{D4_schematic}; two more spheres would be required to contain them. Similarly more double-cones would be required to contain all the meridians of the globe. A total of $2 \times 3 \times 8 = 48$ half-branches = 24 branches are necessary to close the system, that is, to include all branches that emanate from all bifurcations encountered by branches created at FP1. (Other phase changes in $z$ lead to a continuous infinity of branches.)  It is not possible (or we have not been able) to depict the entire scenario in a single diagram. However, we again emphasize that figure \ref{small_FP_xz} can be understood without recourse to this large number of other symmetry-related branches.

\par We have depicted FP4 as connecting two versions of FP1 related by a shift of $\Delta z=1.25$, while passing continuously through a transcritical bifurcation at FP2. However, there exists another construction of this scenario: the transcritical bifurcation could be broken apart in such a way that rather than traversing FP2 smoothly, FP4 enters the transcritical bifurcation at FP2 but then exits at FP2$^\prime\equiv \tau(1.25,0)$FP2. On figure \ref{small_FP_xz}, the second row would not contain a repeated version of figure \ref{small_FP_xz}(f), but instead a shifted version of it. Figures \ref{small_FP_xz}(n,m,l,k) would then also be shifted, with the result that figure \ref{small_FP_xz}(k) would be identical to figure \ref{small_FP_xz}(j) instead of a shifted version of it. In the schematic figure \ref{D3_schematic}, the left cone and right cone would each be reflected in such a way as to separate the two vertices and to join the two bases. In the schematic figure \ref{D4_schematic}, the equator would be cut open, while the north pole and a rotated south pole would be joined. In summary, the FP4 branches start and end in states with a shift of $\Delta z=1.25$ between them, but this can happen either by connecting FP1 and FP1$^\prime$ while passing continuously through FP2, as discussed in sections \ref{tour}-\ref{D3_nf}, or alternatively, by connecting FP2 and FP2$^\prime$ while passing continuously through FP1. Yet another alternative point of view is to double the FP4 branch, passing continuously through FP1, FP2, FP1$^\prime$, FP2$^{\prime}$, FP1 without any phase jumps.

\par \cite{dangelmayr1986steady} determined the normal form for the occurrence of bifurcations to periodic patterns of two different wavenumbers and its unfolding. The equations for the $D_3$-$D_4$ mode interaction predict a number of the features that we observe for our branch FP4 connecting the four-roll and three-roll branches FP1 and FP2, such as the existence of two qualitatively different connecting branches (like our roll-disappearing and roll-merging branches), a nearby saddle-node bifurcation on one of them (like that of figure \ref{small_FP_xz}(g)), and a Hopf bifurcation giving rise to temporally periodic solutions (such as the PO1 branch to be discussed in \S \ref{sec:PO1-PO2}). Some of the features of our scenario are not present in the normal form, in particular the subcriticality of the bifurcations from FP1 to FP4 and the possibly related two additional saddle-node bifurcations of FP4. Another feature which is not mentioned in \cite{dangelmayr1986steady} is the involvement of the two $L_z/8$ phase-shifted versions of FP1. The normal form predicts the stable and unstable eigenvalues of the solution branches, such as those discussed in the next subsection.

\subsection{Stability} 
\label{equilibria_stability}
\par We start by discussing the stability of the FP4 state, which changes multiple times along its branch. Bifurcating subcritically from FP1, the upper (roll-merging) branch of FP4 is initially unstable and inherits the four unstable eigendirections of FP1 at $Ra=13383.9$ (two of which are the $y$-dependent modes that give rise to FP3). Three of these four positive eigenvalues are successively stabilized along the upper branch by undergoing tertiary bifurcations to quaternary states that are not discussed in this work. The last positive eigenvalue is stabilized after undergoing a saddle-node bifurcation at $Ra=8255$. This stability is short lived, however, ending when FP4 undergoes a Hopf bifurcation at $Ra=9980$ to a periodic orbit to be discussed in the next section. The FP4 branch undergoes two more saddle-node bifurcations, at $Ra=11437$ and then at $Ra=10020$, before it finally terminates on FP1$^\prime$, a translated version of the FP1 branch that also has four unstable eigendirections. The connector branch FP4 is thus stable over the interval $8255<Ra<9980$, along with the four-roll branch FP1. However, it is not surprising that the FP4 branch was overlooked by \citet{Gao2013}, since it is unstable over most of its range of existence and its bifurcation occurs from a point at which FP1 is unstable.

\par As mentioned in \S \ref{primary_bif}, in domain $[L_x, L_y, L_z] = [1, 1, 10]$, branch FP1 is stable at onset while FP2 is unstable. Both FP1 and FP2 exist in domain $[1, 0, 10]$ as well. After the bifurcation to FP1, the conductive state acquires two unstable $y$-independent eigenmodes. These are inherited by FP2 at onset, and so FP2 is also unstable when computed in domain $[1, 0, 10]$. Concerning the stability of FP1, since the bifurcation to FP3 breaks $y$-translation symmetry, it does not occur in domain $[1,0,10]$ and FP1 remains stable until the bifurcation to FP4 (in \S \ref{simultan_bif_small}) at $Ra=13383.9$. Regarding FP4, its range of stability does not change if computed in domain $[1,0,10]$, since the unstable part of the FP4 branch always has at least one unstable $y$-independent eigenmode. Due to their four-fold symmetry, FP1 and FP3 can exist in domain $[1, 1, 2.5]$, in which their existence and stability ranges are the same as in $[1, 1, 10]$. These ranges are summarized and compared in table \ref{summary_small}. Since every zero-crossing of an eigenvalue or of its real part is accompanied by a bifurcation, there necessarily exist many more branches that we have not computed, for example along the FP4 branch.

\section{Periodic orbits}
\label{part1_PO}
\begin{figure}
    \centering
    \includegraphics[width=\columnwidth]{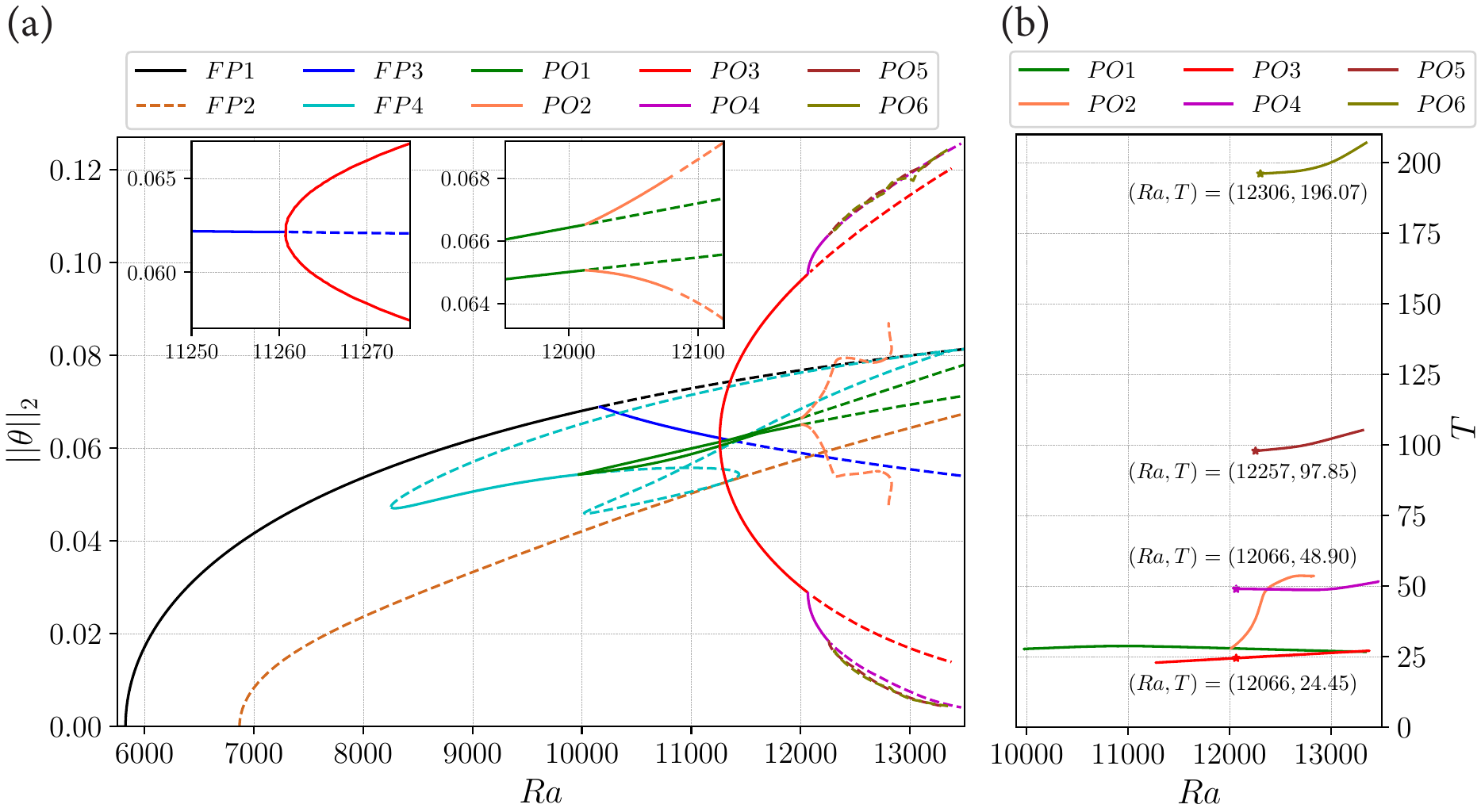}
    \captionsetup{font={footnotesize}}
    \captionsetup{width=13.5cm}
    \captionsetup{format=plain, justification=justified}
    \caption{\label{smalldomain_BDall}(a) Bifurcation diagram containing four fixed points (FPs) and six periodic orbits (POs), in domain $[L_x, L_y, L_z] = [1, 1, 10]$. For each periodic orbit, two curves (maximum and minimum of $\lvert\lvert \theta \lvert\lvert_2$ along an orbit) are shown. PO1 bifurcates from FP4 at $Ra=9980$, PO2 bifurcates from PO1 at $Ra=12013$ and undergoes a saddle-node bifurcation at $Ra=12832$. PO3 bifurcates from FP3 at $Ra=11261$, followed by a period-doubling cascade creating PO4-6 at $Ra = 12066$, $12257$ and $12306$. The apparent lack of smoothness in the curves representing PO6 at $Ra\approx 13000$ corresponds to the overtaking of one temporal maximum of $\lvert\lvert \theta \lvert\lvert_2$ by another as $Ra$ is varied. The two insets zoom in on the Rayleigh number range where PO2 and PO3 bifurcate. Stable and unstable branches are represented by solid and dashed curves, respectively. The stability ranges shown for PO3-6 are those for domain $ [1, 1, 2.5]$; in domain $ [1, 1, 10]$, PO3 is unstable above $Ra=11700$ and so PO4-PO6 are all unstable at onset. (b) The periods of the six periodic orbits in (a), with the same color code. The Rayleigh numbers and periods are listed at each period-doubling bifurcation point, indicated by stars.}
\end{figure}

\par \citet{Gao2013} report that the flow becomes oscillatory at $Ra=11270$ through a Hopf bifurcation from the three-dimensional steady rolls (FP3), and that this is followed by a period-doubling bifurcation at around $12100 < Ra < 12200$. In our simulations of the domain $ [1, 1, 10]$ in the range $9980 < Ra < 11270$, the flow can be either steady or time-periodic, depending on the initial conditions. This suggests that, in addition to the stable steady solution FP3, a limit cycle also exists in this configuration and Rayleigh-number range. We have performed simulations at multiple Rayleigh numbers far from the onset of convection. The time-periodic states have been numerically identified and converged to periodic orbits via the standard Newton shooting approach. These converged time-periodic solutions were subsequently extended in Rayleigh number by parametric continuation. The connections between the periodic orbits and the previously discussed fixed points are shown in the bifurcation diagram in figure \ref{smalldomain_BDall}(a). (As stated previously, other unstable equilibria and periodic orbits exist which we did not investigate or include in figure \ref{smalldomain_BDall}.) This section is devoted to explaining this figure. 

\subsection{Period-doubling cascade: PO3-PO6} 
\par Periodic orbit PO3 arises from FP3 in a supercritical Hopf bifurcation at $Ra=11261$, at which all of the spatial symmetries of FP3 are preserved. This is depicted in the upper left inset of figure \ref{smalldomain_BDall}(a), where the two red branches bifurcating from the FP3 branch correspond to the maximum and minimum of $\lvert\lvert \theta \lvert\lvert_2$ along PO3. PO3 was observed by \cite{Gao2013, Gao2015} and the threshold they reported is $Ra=11270$. Comparing the vorticity isosurfaces of FP3 at $Ra = 11000$ and PO3 at $Ra = 11500$, \cite{Gao2013} show in their figures 10 and 13 that PO3 conserves most of the spatial structure of FP3, with the addition of strands connecting the rolls that appear and disappear periodically.

\par PO3 undergoes a period-doubling bifurcation leading to PO4 at $Ra = 12066$, with further period-doubling bifurcations leading to PO5 and PO6 at $Ra=12257$ and 12306 (very similar to the thresholds $Ra=12068$, $12258$ and $12306$ found by \cite{Gao2015}). The temperature norms $\lvert\lvert \theta \lvert\lvert_2$ of PO3-6 are all close, as is typical for period-doubling cascades. The periods of these limit cycles are shown in figure \ref{smalldomain_BDall}(b), where period-doubling bifurcation points are indicated by stars. We were able to continue all of PO3-6 until at least $Ra=13300$. The dynamics of PO4-6 are very similar to those of PO3 and so we do not show visualizations of them. PO3-6 inherit the four-fold symmetry $[D_4]_{xz}$ of FP3 and hence their spatial and temporal variation all take place within a single roll. These states and the transitions between them are the only phenomena that we report that is not related to competition between three and four rolls. 

\par We refer readers to \citet{Gao2015} for their measurements of the convergence to the Feigenbaum number characterizing the accumulation of period-doubling bifurcations until chaos. Indeed, \citet{Gao2013} observed that the flow in $[L_x, L_y, L_z] = [1, 1, 10]$ becomes irregular following subharmonic oscillations at $Ra=12200$. \citet{Gao2015} estimated that a chaotic regime is reached at around $Ra = 12320$ (in $[L_x, L_y, L_z] = [1, 1, 2.5]$) and our DNS results confirm this. \cite{Gao2015} also discovered and reported five other periodic windows in their table \RomanNumeralCaps{2}, each corresponding to another period-doubling cascade leading to chaotic behavior. Although we have converged and continued these periodic windows, we omit them from the bifurcation diagram to avoid its becoming even more crowded.

\par Domain size has a major effect on the stability of PO3-PO6. When computed in a domain of the size of one wavelength $[1, 1, 2.5]$, PO3 is stable until it is succeeded by PO4, but in the domain $ [1, 1, 10]$, it becomes unstable at $Ra\approx 11700$ by undergoing a large-wavelength instability which breaks the four-fold symmetry. Since this is prior to the period-doubling bifurcation, PO4, PO5, PO6 and the subsequent states resulting from the period-doubling cascade are also unstable in $ [1, 1, 10]$. The stability properties that we have chosen to indicate in \ref{smalldomain_BDall}(a) for PO3-PO6 are those for domain $ [1, 1, 2.5]$. We have summarized the range of existence and stability of PO3-PO6 in both three-dimensional domains in table \ref{summary_small}. Because FP3 is a 3D state, PO3-PO6 do not exist in the two-dimensional domain $[1, 0, 10]$. 

\subsection{Wavelength-changing periodic orbits: PO1-PO2} 
\label{sec:PO1-PO2}
\begin{figure}
    \centering
    \includegraphics[width=\columnwidth]{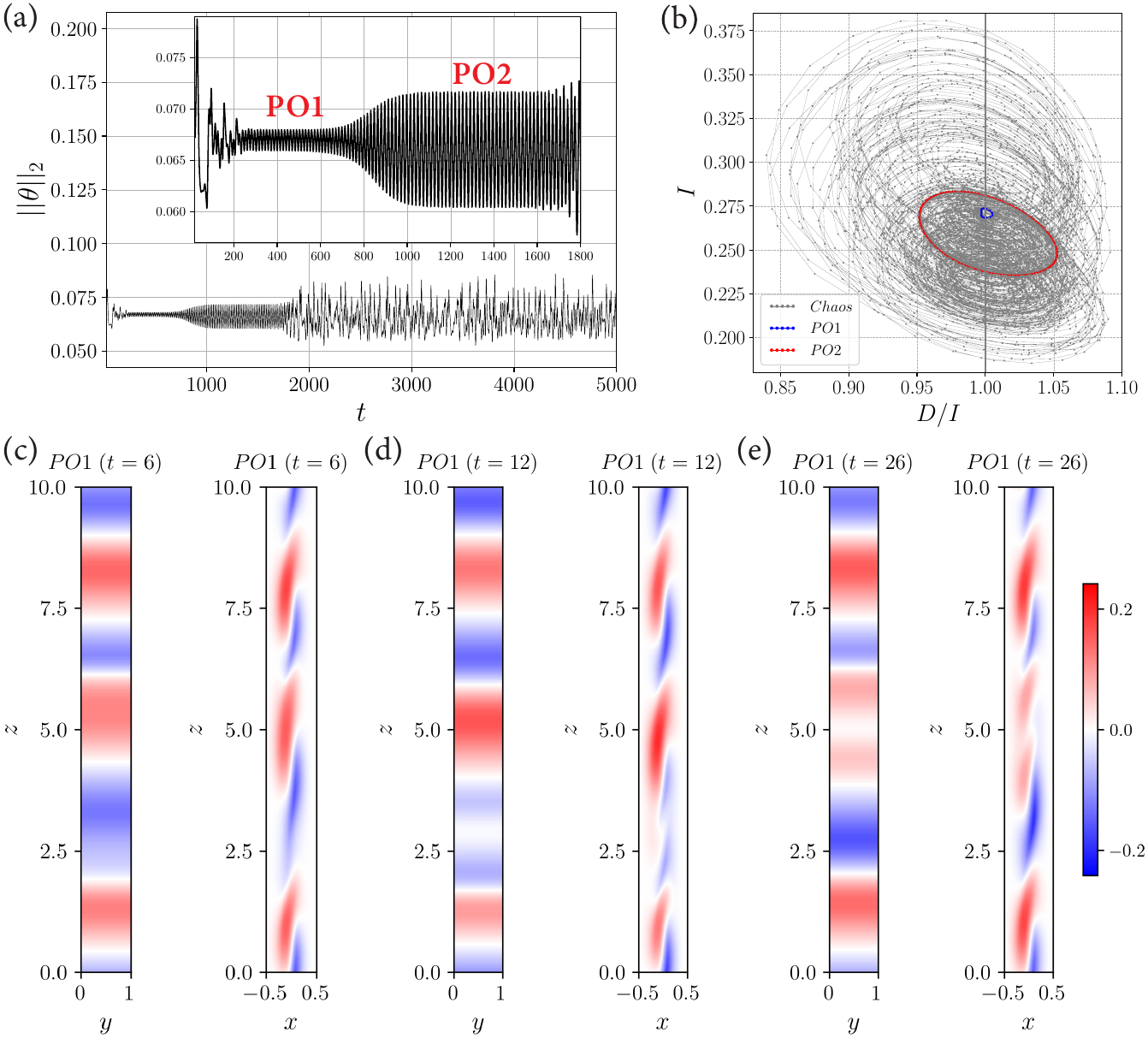}
    \captionsetup{font={footnotesize}}
    \captionsetup{width=13.5cm}
    \captionsetup{format=plain, justification=justified}
    \caption{\label{PO1_PO2} Direct numerical simulations at $Ra=12200$. (a) Time series initialized by random initial conditions. The trajectory passes through two unstable time-periodic flows emphasized in the inset, PO1 ($300<t<650$) with period $T=27.6$, and PO2 ($1000<t<1600$) with period $T=34.5$. (b) Projection of the instantaneous flow fields, separated by $\Delta t=1$, of the chaotic dynamics and of the two periodic orbits onto the thermal energy input ($I$) and the viscous dissipation over energy input ($D/I$). (c-e) Flow structures of PO1 visualized via the temperature field on the $y$-$z$ (at $x=0$) and $x$-$z$ (at $y=0.5$) planes. During the cycle, the longest of the three rolls lengthens and begins to fragment and then recovers. The flow structures of PO2 are shown in figure \ref{PO2}.}
\end{figure}

\begin{figure}
    \centering
    \includegraphics[width=\columnwidth]{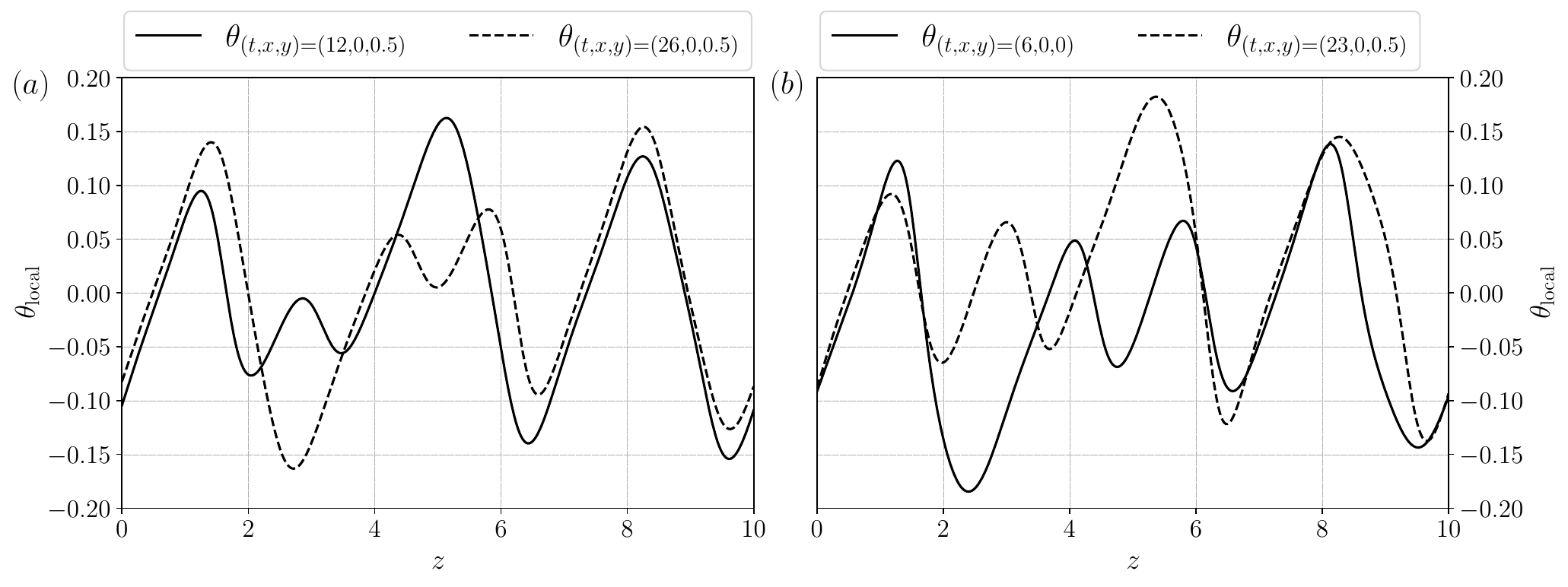}
    \captionsetup{font={footnotesize}}
    \captionsetup{width=13.5cm}
    \captionsetup{format=plain, justification=justified}
    \caption{\label{PO12-series} Temperature profile $\theta_{\rm local}$ at fixed $x$, $y$ and along $z\in[0,10]$ for two instants of PO1 (period $T=27.6$, a) and PO2 (period $T=34.5$, b) at $Ra=12200$. The time interval between two instants for both PO1 and PO2 is approximately half of the corresponding period. For PO1 in (a) the same $x$ and $y$ location are used while for PO2 in (b) the $y$-locations at which the measurements are taken differ by $L_y/2$. The instantaneous temperature fields at $t=12$ and $t=26$ for PO1 as well as at $t=6$ and $t=23$ for PO2 are shown in figures \ref{PO1_PO2} and \ref{PO2}.}
\end{figure}

\begin{figure}
    \centering
    \includegraphics[width=\columnwidth]{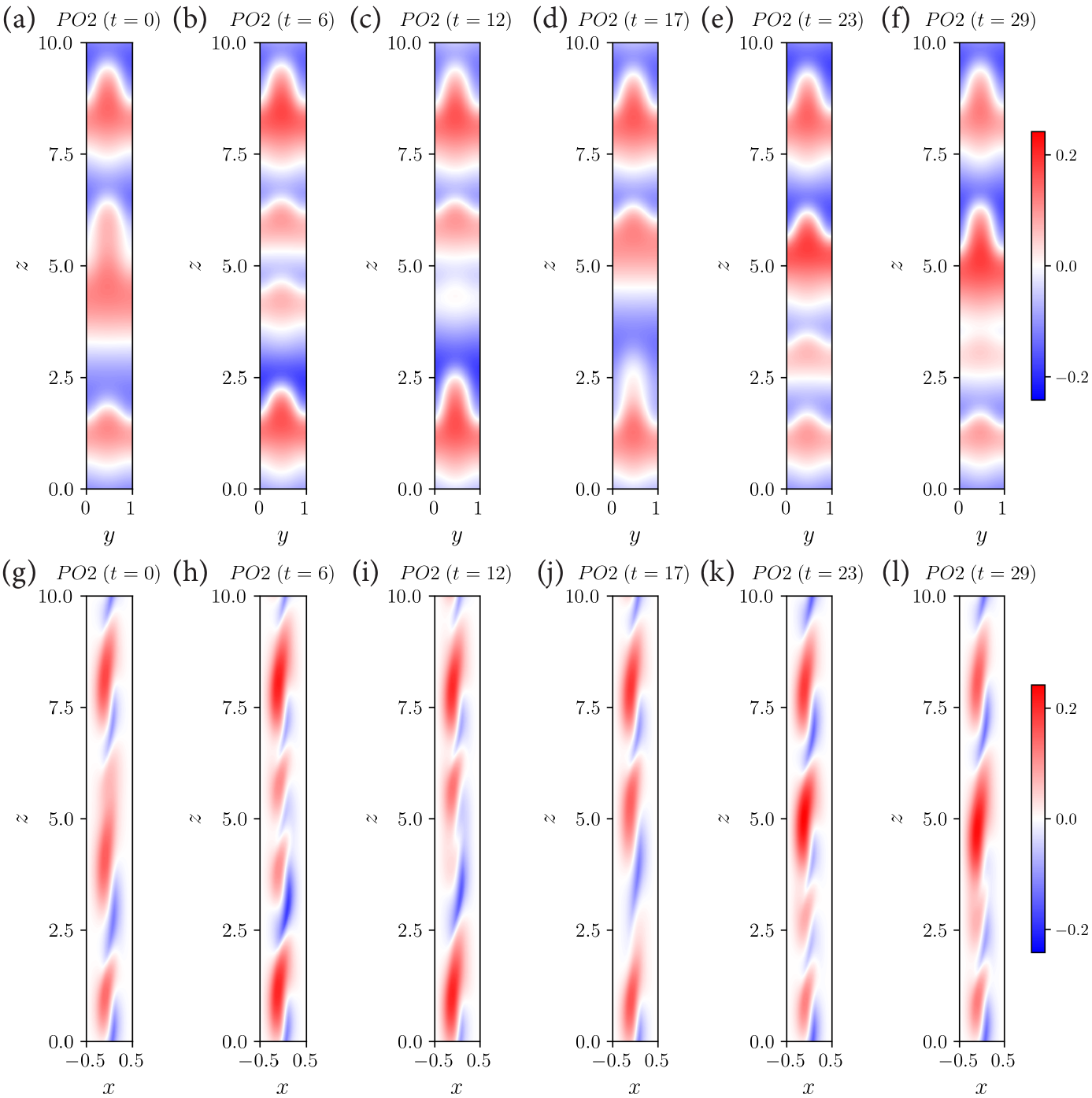}
    \captionsetup{font={footnotesize}}
    \captionsetup{width=13.5cm}
    \captionsetup{format=plain, justification=justified}
    \caption{\label{PO2} Three-dimensional periodic orbit PO2 at $Ra=12200$ (with period $T=34.5$) via two complementary visualisations of the temperature field. (a-f) $y$-$z$ plane at $x=0$ and (g-l) $x$-$z$ plane at $y=0.5$.}
\end{figure}

\par For $Ra>12200$ in domain $ [1, 1, 10]$, \citet{Gao2013} observed that numerical simulations started from a random initial condition settled down to three rolls, while four rolls were found intermittently. They also observed hysteresis in a simulation at $Ra=12000$, initiated from an instantaneous three-roll flow field at $Ra = 12200$ and which finally settled down to a periodic orbit with three rolls. We have carried out a direct numerical simulation at nearly the same parameter values, illustrated in figure \ref{PO1_PO2}(a), which shows long intervals of two types of time-periodic behavior, which we call PO1 and PO2. The simulation is started from a random initial condition. The initial chaotic behavior for $t \lesssim 250$ is succeeded by the weakly unstable PO1 ($300 \lesssim t \lesssim 650$). Afterwards, PO2 is visited ($1000 \lesssim t \lesssim 1600$) before a transition via subcritical period-doubling to a fully chaotic state ($t \gtrsim 1800$), which then persists without relaminarization. A phase portrait of this DNS is shown in figure \ref{PO1_PO2}(b) where instantaneous flow fields are represented by grey dots in the $(D/I, I)$ plane, where $D$ is the dissipation due to viscosity and $I$ is the thermal energy input due to buoyancy. It can be seen that the two periodic orbits, with relatively low input/dissipation, are surrounded by the fully chaotic dynamics in this projection.

\par Continuing PO1 backwards in Rayleigh number, we find that PO1 is created via a supercritical reflection-symmetry-breaking Hopf bifurcation from FP4 at $Ra=9980$ where FP4 is stable, so that PO1 is stable at onset. The complex conjugate neutral eigenvector pair of FP4 is anti-$xz$-reflection symmetric, so PO1 has the spatial symmetry group $S_{PO_1} \equiv \braket{\pi_y, \tau(\Delta y,0)} \simeq [O(2)]_y$, and the spatio-temporal symmetry
\begin{align}
    (u,v,w,\theta)(x,y,z_0+z,t+T/2) = (-u,v,-w,-\theta)(-x,y,z_0-z,t),
\end{align}
where here $z_0 \approx 4$ and $T/2 \approx 14$; compare figures \ref{PO1_PO2}(d) and (e) as well as figure \ref{PO12-series}(a).

\par PO1 arises from FP4 at a Rayleigh number at which it has three unequal rolls; its temporal dynamics consist of periodic lengthening and near-fragmentation of the longest roll. Figure \ref{PO12-series} shows temperature profiles in $z$ for PO1 (and PO2), at fixed $x$ and $y$ locations, and at two instants separated in time by approximately a half-period. In these profiles, two successive zero-crossings ($\theta_{\rm local}(z)=0$) correspond to one roll. Figure \ref{PO12-series}(a) shows a close, but unsuccessful approach to roll creation in PO1 at $z\approx3$ ($t=12$) and then again at $z\approx5$ ($t=26$).

\par PO1 loses stability in a supercritical pitchfork bifurcation at $Ra=12013$ (emphasized in the upper right inset of figure \ref{smalldomain_BDall}(a)), leading to the creation of PO2, in which $y$-translation symmetry is broken but $y$-reflection symmetry is retained: $S_{PO_2} \equiv \braket{\pi_y} \simeq [Z_2]_y$. Unlike PO1, in PO2 the central roll succeeds in splitting, as can be seen in figures \ref{PO2}(b) and (h)) as well as by the zero-crossings in figure \ref{PO12-series}(b), but only temporarily (figures \ref{PO2}(b), (e), (h) and (k)). PO2 has the spatio-temporal symmetry 
\begin{align}
    (u,v,w,\theta)(x,y,z_0+z,t+T/2) = (-u,v,-w,-\theta)(-x,y+0.5,z_0-z,t),
    \label{half_domain_shift}
\end{align}
where $z_0$ takes the value $z_0 \approx 4$ and $T/2 \approx 17$. (The wavy structure of PO2 leads to the requirement that a half-domain shift in $y$ be included in \eqref{half_domain_shift}. The combination of $\tau(0.5,0)$ and $\pi_{xz}$ relates the red upward-facing ``tongues'' centered at $y=0.5$ to the blue downward-facing ``tongues'' centered at $y=0$.) 

\par PO2 undergoes a saddle-node bifurcation at $Ra=12832$ and we followed this PO2 branch until $Ra=12804$, shortly after the saddle-node bifurcation. PO2 undergoes a secondary Hopf bifurcation (also called Neimark-Sacker or torus bifurcation) at $Ra=12082$, at which a complex conjugate pair of eigenvalues crosses the real axis, which generally leads to a torus that we will not discuss in the present work. PO2 is thus stable only over a very small range of Rayleigh number ($12013 < Ra < 12082$), as can be seen in the upper right inset of figure \ref{smalldomain_BDall}(a). It is therefore not surprising that PO2 was not observed by \cite{Gao2013}. The period of PO1 remains nearly constant over a wide range of $Ra$ while that of PO2 increases with $Ra$ and then stays almost constant. 

\par PO1 and PO2 capture the oscillatory dynamics of convection rolls. PO1 has three non-uniform rolls of fluctuating size and intensity (figures \ref{PO1_PO2}(c-e)). Although some rolls stretch and become quite weak, they never actually split anywhere along the branch. For PO2, the number of rolls varies between three (figures \ref{PO2}(a, c, d, f) along with (g, i, j, l)) and four (figures \ref{PO2}(b, e) along with (h, k)). We suggest that the intermittency and hysteresis observed by \citet{Gao2013} are a manifestation of visits to the coexisting unstable periodic orbits PO1 and PO2.

\par Since PO1 is $y$-independent, it can also exist in domain $[L_x, L_y, L_z] = [1, 0, 10]$. While the existence ranges of PO1 in domains $[1, 0, 10]$ and $[1, 1, 10]$ are the same, their stabilities differ: in $[1, 0, 10]$, the bifurcation to $y$-dependent PO2 does not occur, and so PO1 remains linearly stable at least until $Ra=14000$, the upper limit at which we stopped the continuation. \citet{Gao2013} also observed this oscillating three-roll flow in the two-dimensional domain $[1, 0, 10]$, for $13500<Ra<15300$. PO1 and PO2 cannot exist in domain [1,1,2.5] and PO2 cannot exist in domain [1,0,10]. The existence and stability intervals that we have computed for all of these flows are stated and compared with those of \cite{Gao2013, Gao2015} in table \ref{summary_small}.

\section{Discussion and conclusions}
\begin{table}
    \centering
    \begin{tabular}{cccc}
        Flow  & Present results & Present results & \cite{Gao2013,Gao2015} 
         \\   & existence       & stability       & \\
        \hline
        \textbf{Domain [1, 1, 10]} \\
        base flow  & $0 \leq Ra$ & $0\leq Ra<5826$ &  $0\leq Ra < 5800$ \\
        FP1	(4 2D rolls) & $ ~5826 <Ra < 14000$ &  $~5826 < Ra < 10166$ & $5800 < Ra < 10100$ \\
        FP2	(3 2D rolls) & $~6869 < Ra < 14000$  & unstable  & - \\
	FP3 (4 3D rolls) & $10166<Ra<14000$  &  $10166 < Ra < 11261$ &$10100 < Ra < 11270$ \\
        FP4 (connector) &  $~8255 < Ra < 13384$ &  $~8255 < Ra < 9980$ & - \\
        PO1 & $~9980 < Ra < 14000$ & $9980 < Ra < 12013$  & $Ra=12000$ \\
        PO2 & $12013 < Ra < 12832$ & $12013 < Ra < 12082$  & $12200<Ra\lesssim 13000$ \\ 
        PO3 & $11261 < Ra < 13300$ & $11261 < Ra \lesssim 11700$ & $11270 < Ra < 12000$ \\
        PO4 &  $12066 < Ra < 13300$  & unstable & $12100 < Ra < 12200$ \\   
        PO5  &  $12257 < Ra < 13300$  & unstable & - \\
        PO6  &  $12306 < Ra < 13300$  & unstable & - \\
        \hline
        \textbf{Domain [1, 1, 2.5]} \\
        base flow  & $0 \leq Ra$ & $0\leq Ra<5826$ &  $0\leq Ra < 5800$ \\
        FP1 (1 2D roll) & $ ~5826 <Ra < 14000$ &  $~5826 < Ra < 10166$ & $5800 < Ra < 10100$ \\
        FP3 (1 3D roll) & $10166<Ra<14000$  &  $10166 < Ra < 11261$ &$10100 < Ra < 11270$ \\
        PO3 & $11261 < Ra < 13300$  & $11261 < Ra < 12066$ & $11270 < Ra < 12068$ \\
        PO4 & $12066 < Ra < 13300$  & $12066 < Ra < 12257$ & $12068 < Ra < 12258$ \\  
        PO5 & $12257 < Ra < 13300$  & $12257 < Ra < 12306$ & $12258 < Ra < 12306$ \\ 
        PO6 & $12306 < Ra < 13300$  & $12306 < Ra < 12316$ & $12306 < Ra < 12317$ \\  
        \hline
        \textbf{Domain [1, 0, 10]} \\
        base flow & $0 \leq Ra$ & $0\leq Ra<5826$ & $0\leq Ra<5708$ \\
        FP1 (4 2D rolls) & $ ~5826 <Ra< 14000$ & $~5826 < Ra < 13384$ & $5708<Ra<13000$ \\
        FP2 (3 2D rolls) & $ ~6869 <Ra< 14000$ & unstable & - \\
        FP4 (connector) & $~8255 < Ra < 13384$ & $~8255 < Ra < 9980$ & - \\
        PO1 & $~9980 < Ra < 14000$ & $9980 < Ra < 14000$ & $13500<Ra<15300$ \\
    \hline
    \end{tabular}
    \captionsetup{font={footnotesize}}
    \captionsetup{width=13.5cm}
    \captionsetup{format=plain, justification=justified}
    \caption{Summary of bifurcation sequence and comparison with the literature, including all fixed points (FPs) and periodic orbits (POs) mentioned in this paper. All of the states exist in domain $[L_x, L_y, L_z] = [1, 1, 10]$, while only some exist in smaller domains $[1, 1, 2.5]$ and $[1, 0, 10]$. For each of the three domains, we summarize the ranges of existence and stability for all states that we have computed. States existing in two domains may be stable in the smaller domain but unstable in the larger domain.  When upper limits are listed as $14000$ or $13300$, these numbers are not the end of the branch, but where we stopped the numerical continuation. The Rayleigh number ranges given in the last column correspond to those reported by \citet{Gao2013, Gao2015} and do not necessarily strictly correspond to existence or stability ranges. Ranges not reported are indicated by “-”.}
    \label{summary_small}
\end{table}

\par Vertical convection supports a large variety of flow patterns and thus can serve as a paradigm for non-linear pattern formation in driven dissipative out-of-equilibrium systems. In this work, we have investigated thermal convection between two vertical plates held at different temperatures via both numerical simulation and continuation. We have computed the stable and unstable invariant solutions of the fully non-linear three-dimensional Oberbeck–Boussinesq equations leading to the spatio-temporal complex convection patterns observed in experiments and simulations, far beyond the onset of convection.

\par We have also discovered previously unknown fixed points and periodic orbits. For these and the previously known solutions, we have identified the bifurcations responsible for their generation and termination, as well as their stabilization and destabilization. Summarizing the bifurcations and regimes in the computational domains $[L_x, L_y, L_z] = [1, 1, 10]$, [1, 1, 2.5] and [1, 0, 10], we compare in table \ref{summary_small} the results from \citet{Gao2013, Gao2015} with those that we have obtained by computing unstable states and their bifurcations. Good quantitative agreement is achieved for those states which are observed in both studies. We note that all of the solution branches that we have found are connected by one or more bifurcations to the laminar branch.

\par Despite the complexity of the bifurcation diagrams shown in figures \ref{small_FP} and \ref{smalldomain_BDall}, almost all of the dynamics of this system result from one simple physical phenomenon: the competition between three and four rolls. (The exception is the steady three-dimensional state FP3 and the subsequent period-doubling sequence PO3-PO6, whose dynamics involve a single roll.) This competition takes different forms for the steady and time-dependent states. 

\par For steady states, the most basic scenario in pattern formation is essentially one-dimensional and consists of a sequence of primary bifurcations from a featureless state to branches with different numbers of regularly spaced rolls, cells, or waves. Of these, only the first to bifurcate is stable at onset; primary branches often undergo secondary instabilities to mixed-mode branches which transfer stability between the different branches \cite[e.g.,][]{knobloch1983convective}. In an idealized context, this is the Eckhaus instability \cite[e.g.,][]{eckhaus1965studies, tuckerman1990bifurcation}. A mathematical framework which covers wavelength competition is that of mode interaction \citep{dangelmayr1986steady}. 

\par Our hydrodynamic configuration involving the four-roll branch FP1, the three-roll branch FP2 and the connector branch FP4 presents a complicated version of this basic scenario. The FP1 branch with four equally spaced rolls has $D_4$ symmetry and hence gives rise to two sets of mixed-mode branches FP4. These two sets are associated with two qualitatively different paths for passing from four to three rolls: the merging of two rolls and the disappearance of a roll. (This is kinematically possible because these are co-rotating rolls, rather than the counter-rotating rolls of the standard Rayleigh-B\'enard configuration.) Indeed, dual sets of branches are a typical feature of bifurcation in the presence of $D_4$ symmetry \citep{swift1985bifurcation, Knobloch1986}. The $D_4$ bifurcation scenario is present in many other situations and will be encountered again in our companion paper \citet{Zheng2023part2}, in the more classic two-dimensional context of competition between straight and diagonal orientations \citep{demay1984calcul, tagg1989nonlinear, Chossat1994, bengana2019spirals, Reetz2020b}.

\par A new feature of the $D_4$ scenario seen here is that each FP4 half-branch of disappearing-roll type meets and merges smoothly with an FP4 half-branch of merging-roll type. To the best of our knowledge, this phenomenon has not been previously observed. The FP4 branches which merge do not emanate from the same four-roll branch, but from two branches, FP1 and FP1$^\prime$, that are phase-shifted by a half-roll with respect to one another. The simultaneous existence of branches FP1 and FP1$^\prime$ is in turn due to the fact that FP1 branches of all phases are created by a circle pitchfork bifurcation. At this meeting point, the two types of half-branches also meet the FP2 branch, which contains states with three equal rolls, in a transcritical bifurcation. The phase jump can instead be assigned to the transcritical bifurcation, i.e.\ between three-roll branches FP2 and FP2$^\prime$, rather than to the pitchfork bifurcations from the four-roll branches. A last alternative is to follow the FP4 branch twice, via FP1, FP2, FP1$^\prime$, FP2$^\prime$, FP1 without any phase jumps. The $D_3$ symmetry of the FP2 states governs the details of the transcritical bifurcation. The physical phenomena of roll-merging and roll-disappearance, roll-creation and roll-splitting provide visual illustrations of the group-theoretic $D_4$ and $D_3$ scenarios and of the $D_3$--$D_4$ mode-interaction equations in \cite{dangelmayr1986steady} and \cite{crawford1990period}.

\par Turning now to time-dependent solutions, periodic orbits PO1 and PO2 could be considered to be temporal versions of the variation with Rayleigh number along the connector branch FP4, from which PO1 bifurcates. Figure \ref{PO1_PO2} shows that PO1 contains temporal phases in which one of its three rolls widens or weakens, resembling the precursors to four rolls seen in figure \ref{small_FP_xz} as $Ra$ is varied along the FP4 branch. Although PO1 does not succeed in creating a fourth roll, figure \ref{PO2} shows that in PO2 these events culminate in the periodic formation and destruction of a fourth small and fragile roll. This may or may not be related to the breaking of $y$-translation symmetry from PO1 to PO2; perhaps roll formation or destruction is facilitated when rolls become wavy. The competition between three and four rolls continues to dominate for Rayleigh numbers above 12082 when PO2 is destabilized, since the dynamics beyond this point consist of chaotic three-roll flow with infrequent and irregular bursts of four rolls, as illustrated by \citet{Gao2013}. In the present study in domain $[L_x, L_y, L_z] = [1, 1, 10]$, we report no stable fixed points for $Ra>11261$ and no stable periodic orbits for $Ra>12082$. Nevertheless, we have been able to numerically continue these unstable solutions into the chaotic regime, far from the parameter regime in which they are stable. 

\par Although direct numerical simulations can give access to complex flow dynamics at specific control parameters, numerical continuation organizes these solutions and determines their bifurcation-theoretic origin, by situating them in the context of a bifurcation diagram. Our work bridges the gap between purely DNS-based observations and numerical bifurcation analysis, leading to a better description and understanding of complex convective flows.

\backsection[Acknowledgements]{We thank Sajjad Azimi, Florian Reetz and Omid Ashtari for fruitful exchanges. We thank Patrick Le Qu\'er\'e for sharing his encyclopedic knowledge of vertical convection. We are grateful to Dwight Barkley, Jonathan Dawes, Edgar Knobloch, and Alastair Rucklidge for their insights on symmetry. We are grateful to the anonymous referee who added the alternative construction of the connector branch, and to all of the referees for their very valuable comments.}

\backsection[Funding]{This work was supported by the European Research Council (ERC) under the European Union's Horizon 2020 research and innovation programme (grant no. 865677).}

\backsection[Declaration of interests]{The authors report no conflict of interest.}

\backsection[Author ORCID]{
\newline
Zheng Zheng \href{https://orcid.org/0000-0002-9833-1347}{https://orcid.org/0000-0002-9833-1347};
\newline
Laurette S. Tuckerman \href{https://orcid.org/0000-0001-5893-9238}{https://orcid.org/0000-0001-5893-9238}; 
\newline
Tobias M. Schneider \href{https://orcid.org/0000-0002-8617-8998}{https://orcid.org/0000-0002-8617-8998}.}

\bibliographystyle{jfm}

\end{document}